\documentclass[traditabstract]{aa}

\newcommand{\myemail}{mj.michalowski@gmail.com}
\newcommand{\urltt}[1]{\url{\texttt{#1}}}

\usepackage{graphicx}
\usepackage{txfonts}

\newcommand{\msun}{\mbox{$M_\odot$}}

\newcommand{\micron}{\mbox{$\mu$m}}

\newcommand{\hi}{\sc Hi}
\newcommand{\mhi}{M_{\rm HI}}
\newcommand{\kms}{\mbox{km\,s$^{-1}$}}

\newcommand{\sn}{SN\,2002ap}

\begin{document}

\title{Connection of supernovae  2002ap, 2003gd, 2013ej, and 2019krl in M74 with atomic gas accretion and spiral structure
}

\titlerunning{Gas in the vicinity of SN 2002ap, 2003gd, 2013ej, and 2019kr in M74}
\authorrunning{Micha{\l}owski et al.}
 
\author{Micha{\l}~J.~Micha{\l}owski\inst{\ref{inst:uam},\ref{inst:cal},\ref{fulb}}, 
        Natalia~Gotkiewicz\inst{\ref{inst:umk}},
        Jens~Hjorth\inst{\ref{inst:dark}},
        Peter Kamphuis\inst{\ref{inst:airub}}
        }

\institute{
Astronomical Observatory Institute, Faculty of Physics, Adam Mickiewicz University, ul.~S{\l}oneczna 36, 60-286 Pozna{\'n}, Poland \myemail  \label{inst:uam}
\and
TAPIR, Mailcode 350-17, California Institute of Technology, Pasadena, CA 91125, USA  \label{inst:cal}
\and
Fulbright Senior Award Fellow \label{fulb}
\and
Institute of Physics, Faculty of Physics, Astronomy and Informatics, Nicolaus Copernicus University, Grudzi\c{a}dzka 5, 87-100 Toru\'{n}, Poland\label{inst:umk}
\and
DARK, Niels Bohr Institute, University of Copenhagen, Lyngbyvej 2, 2100 Copenhagen, Denmark\label{inst:dark}
\and
Ruhr-Universit\"at Bochum, Faculty of Physics and Astronomy, Astronomical Institute, 44780 Bochum, Germany\label{inst:airub}
}


\abstract
{Studying the nature of various types of supernovae (SNe) is important for our understanding of stellar evolution.
Observations of atomic and molecular gas in the host galaxies of gamma-ray bursts (GRBs) and SNe have recently been used to learn about the nature of the explosions themselves and the star formation events during which their progenitors were born.
Based on archival data for M74, which previously has not been investigated in the context of SN positions, we report the gas properties in the environment of the broad-lined type Ic (Ic-BL) SN\,2002ap and the type II SNe 2003gd, 2013ej, and 2019krl. 
{The \sn} is located  at  the end of an off-centre, asymmetric, 55\,kpc-long {\hi} extension containing 7.5\% of the total atomic gas in M74, 
interpreted as a signature of external gas accretion.
It is the fourth known case of an explosion of a presumably massive star located close to a concentration of atomic gas 
(after GRBs\,980425, 060505, and SN\,2009bb). 
It is unlikely that all these associations are random (at a $3\sigma$ significance), so 
the case of {\sn} adds to the evidence that the birth of the progenitors of type Ic-BL SNe and GRBs is connected with the accretion of atomic gas from the intergalactic medium.
The {\hi} extension could come from tidally disrupted companions of M74, or be a remnant of a galaxy or a gas cloud that accreted entirely from the intragroup medium. 
The other (type II) SNe in M74  are located at the outside edge of a spiral arm.
This suggests that either their progenitors were born when gas was piling up there or that the SN progenitors moved away from the arm due to their orbital motions.
These type II SNe do not seem to be related to gas accretion.
}


\maketitle

\section{Introduction}

\newlength{\szerkol}

\setlength{\szerkol}{0.5\textwidth}

\begin{figure*}
\begin{center}
\begin{tabular}{ccc}
\hspace{-1.em}\includegraphics[width=1\szerkol,clip,viewport=2 150 575 690]{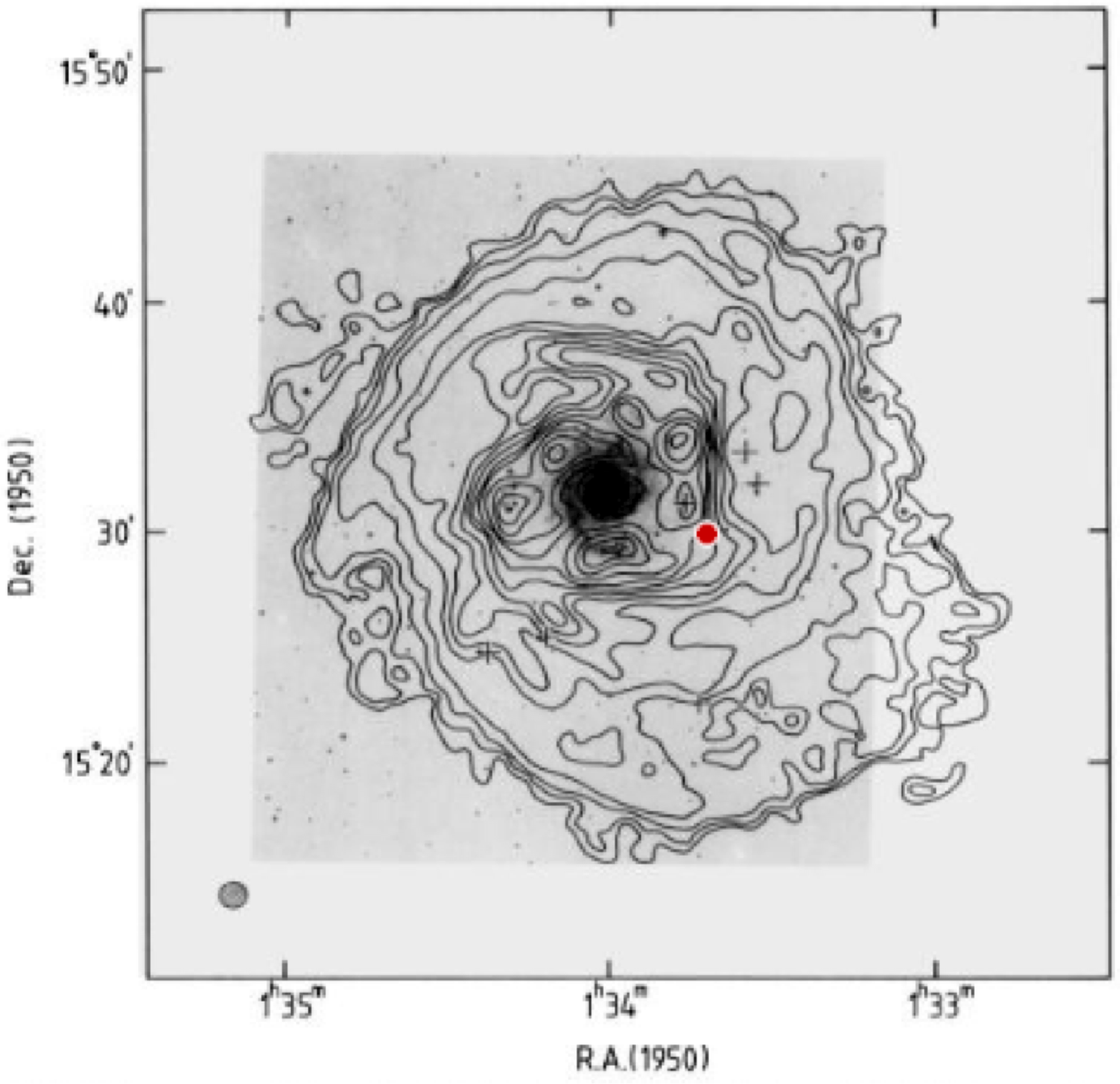} &
\hspace{-1.5em}\includegraphics[width=1\szerkol,clip,viewport=2 150 575 690]{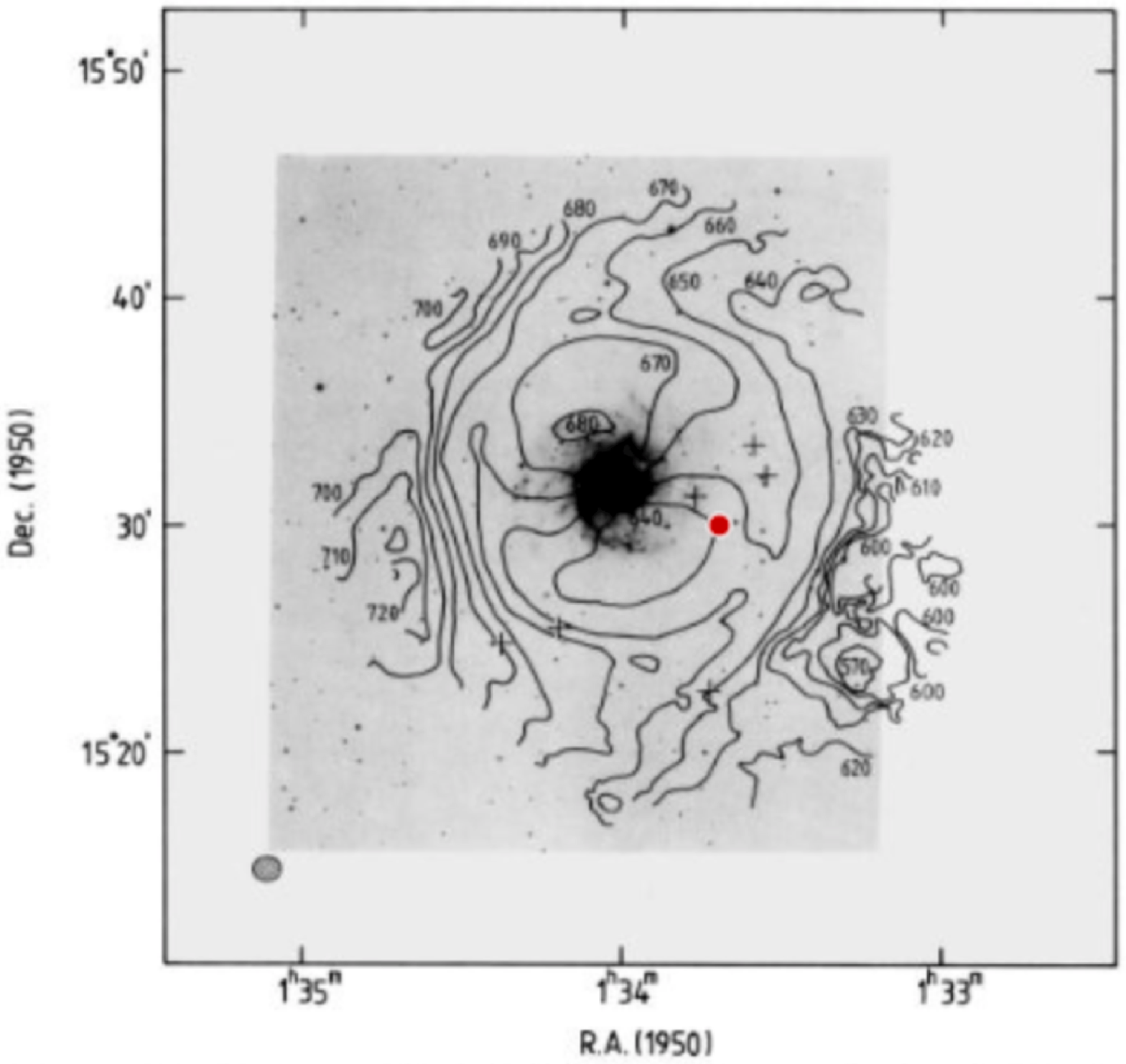}\\

\end{tabular}
\end{center}
\caption{
Low resolution  {\hi} map
(the beam size of $72\arcsec\times62\arcsec$)
superimposed on an optical image of M74.
The position of {\sn} is marked as the red dot
(credit: Kamphuis \& Briggs, 1992, reproduced with permission $\copyright$ ESO).
Left: Zeroth moment map (integrated emission).
Right:
First moment map (velocity fields).
The asymmetric tail with an irregular velocity field is visible at the south-western outskirt of the atomic disc (around the position of $\mbox{R.A.}=1^h 33^m20^s$, $\mbox{Dec.}=15^\circ 25^m$).
North is up and east is to the left.
}
\label{fig:lowres}
\end{figure*}

\setlength{\szerkol}{0.3\textwidth}

\begin{figure*}
\begin{center}
\begin{tabular}{ccc}
\includegraphics[width=\szerkol,clip]{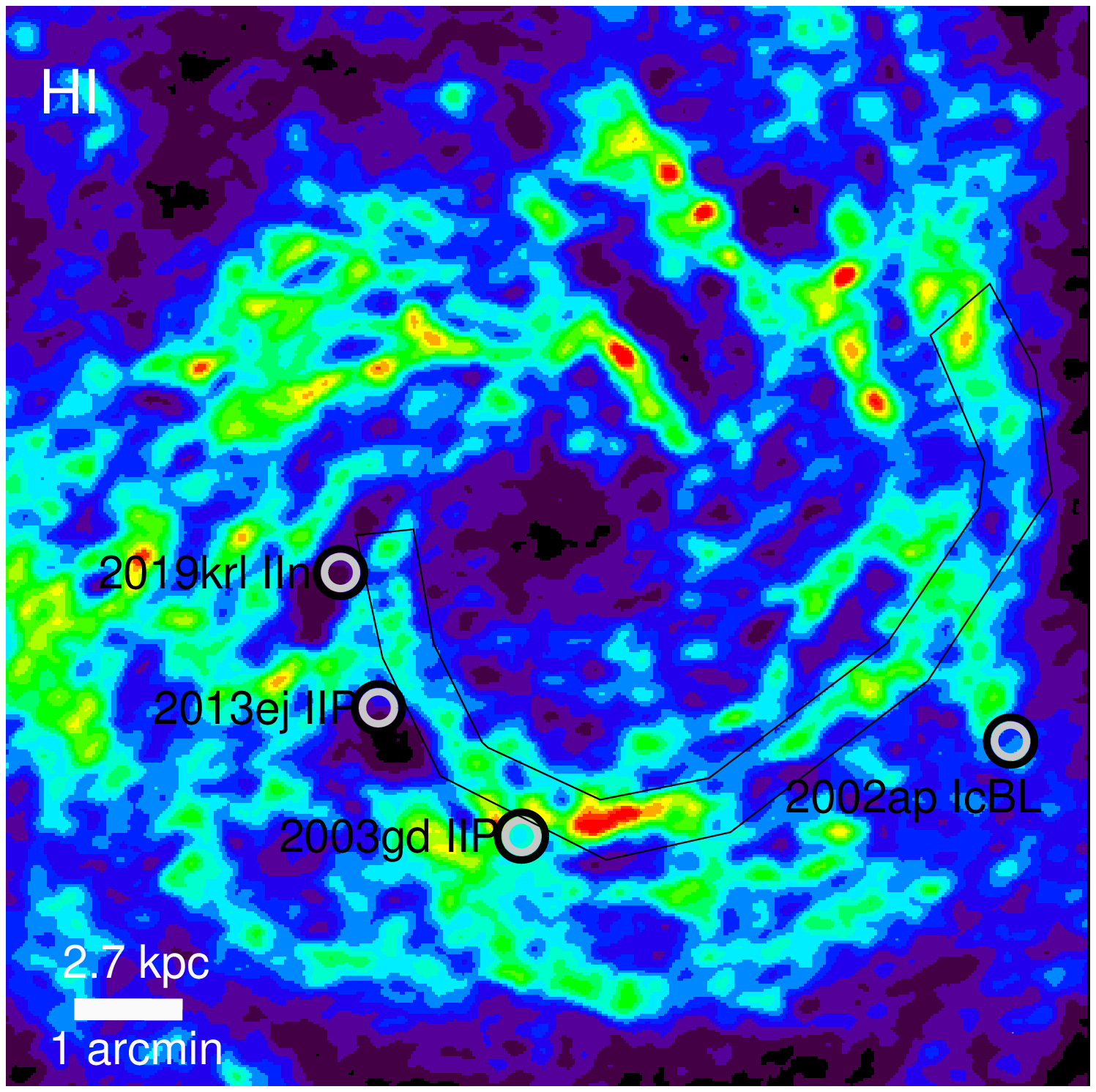} & 
\includegraphics[width=\szerkol,clip]{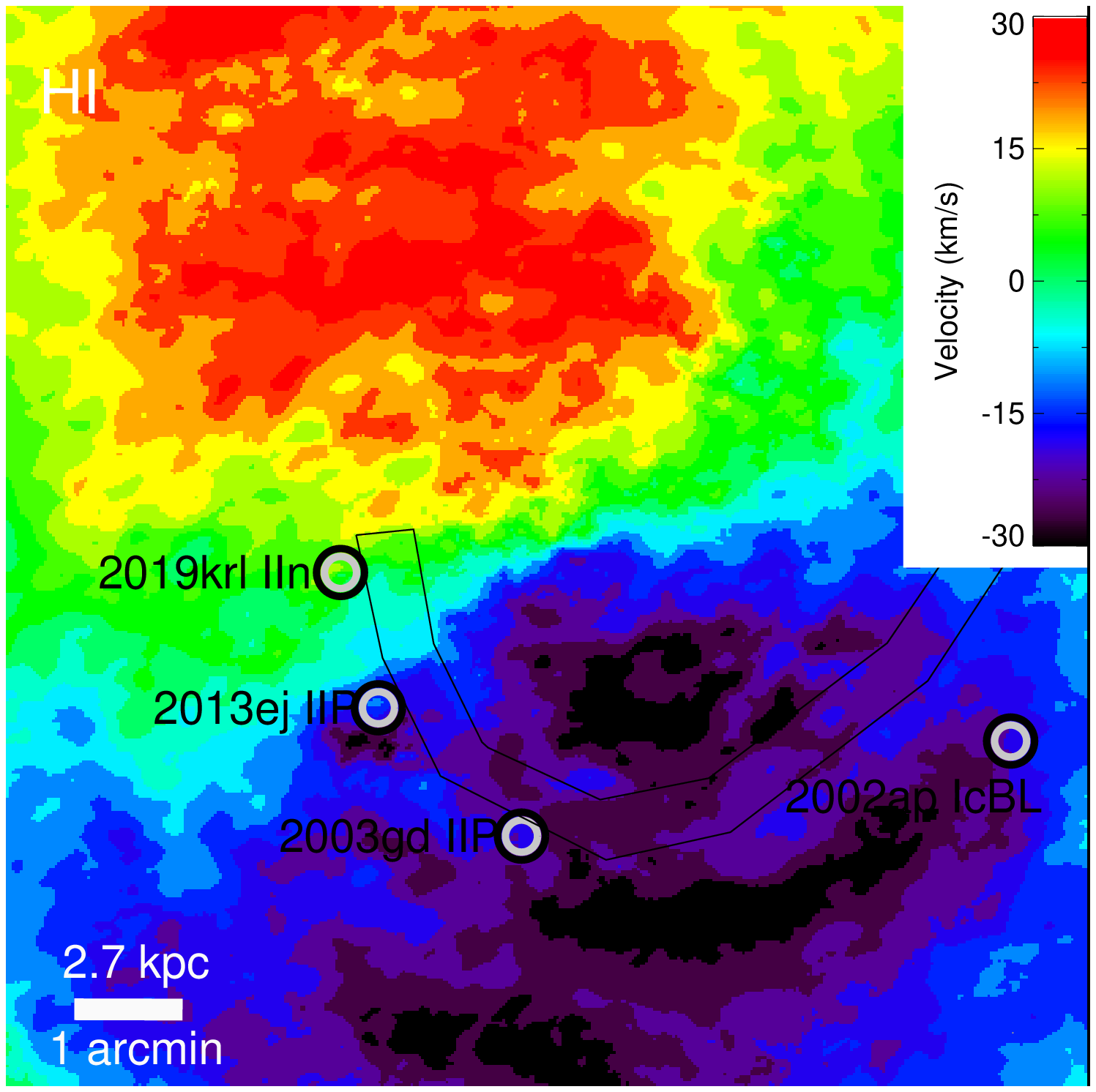} & 
\includegraphics[width=\szerkol,clip]{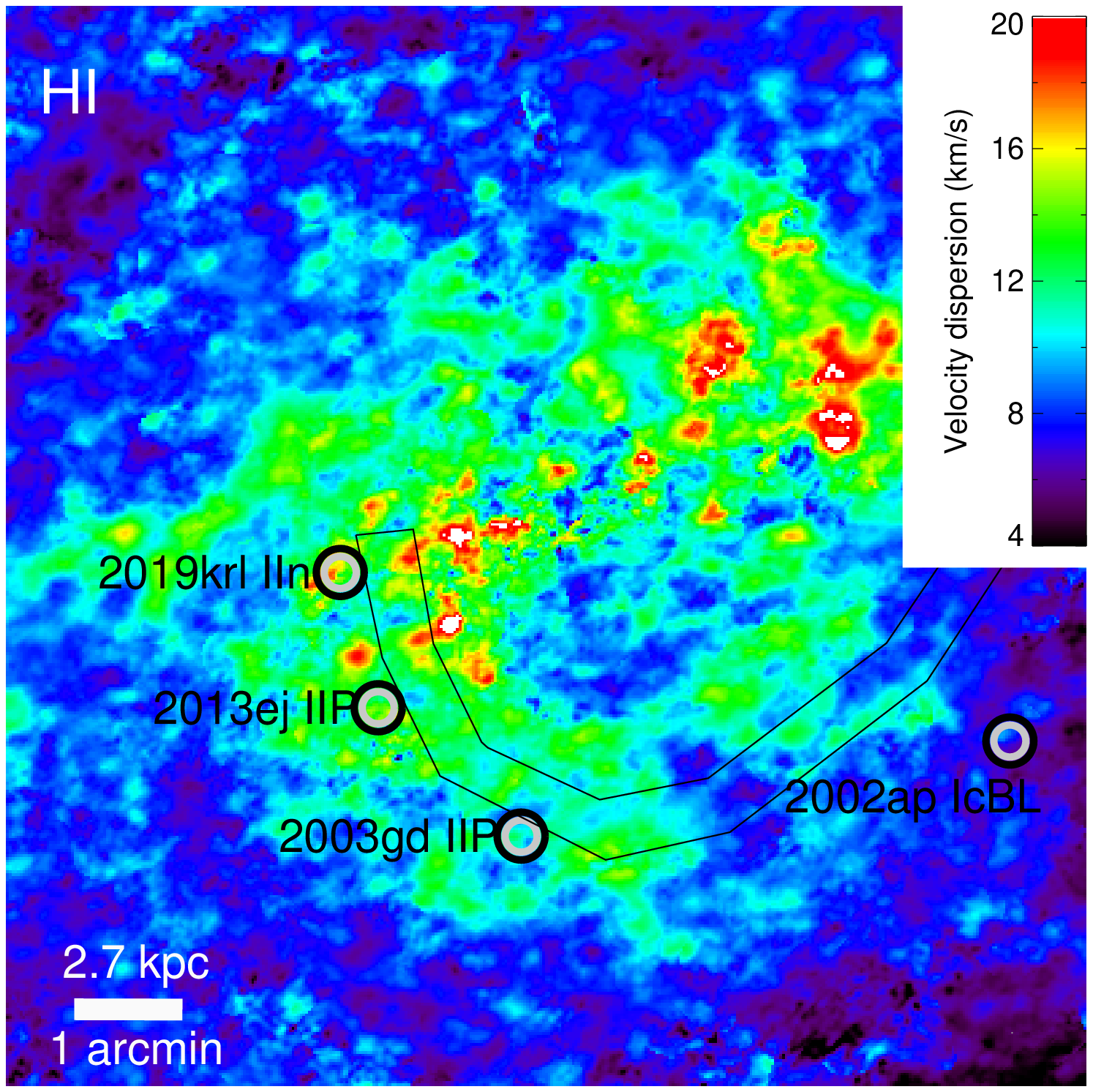} \\
\includegraphics[width=\szerkol,clip]{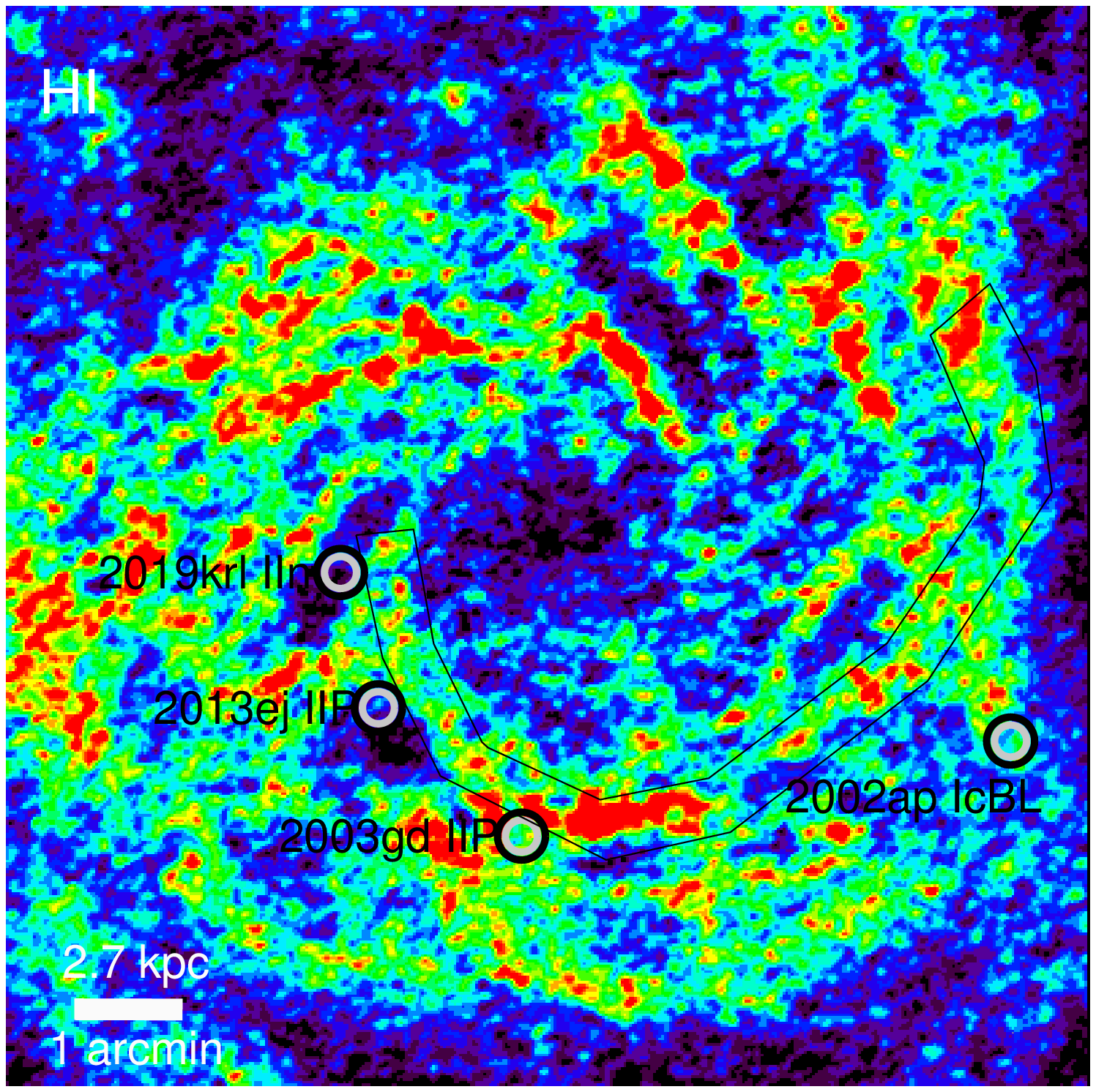} & 
\includegraphics[width=\szerkol,clip]{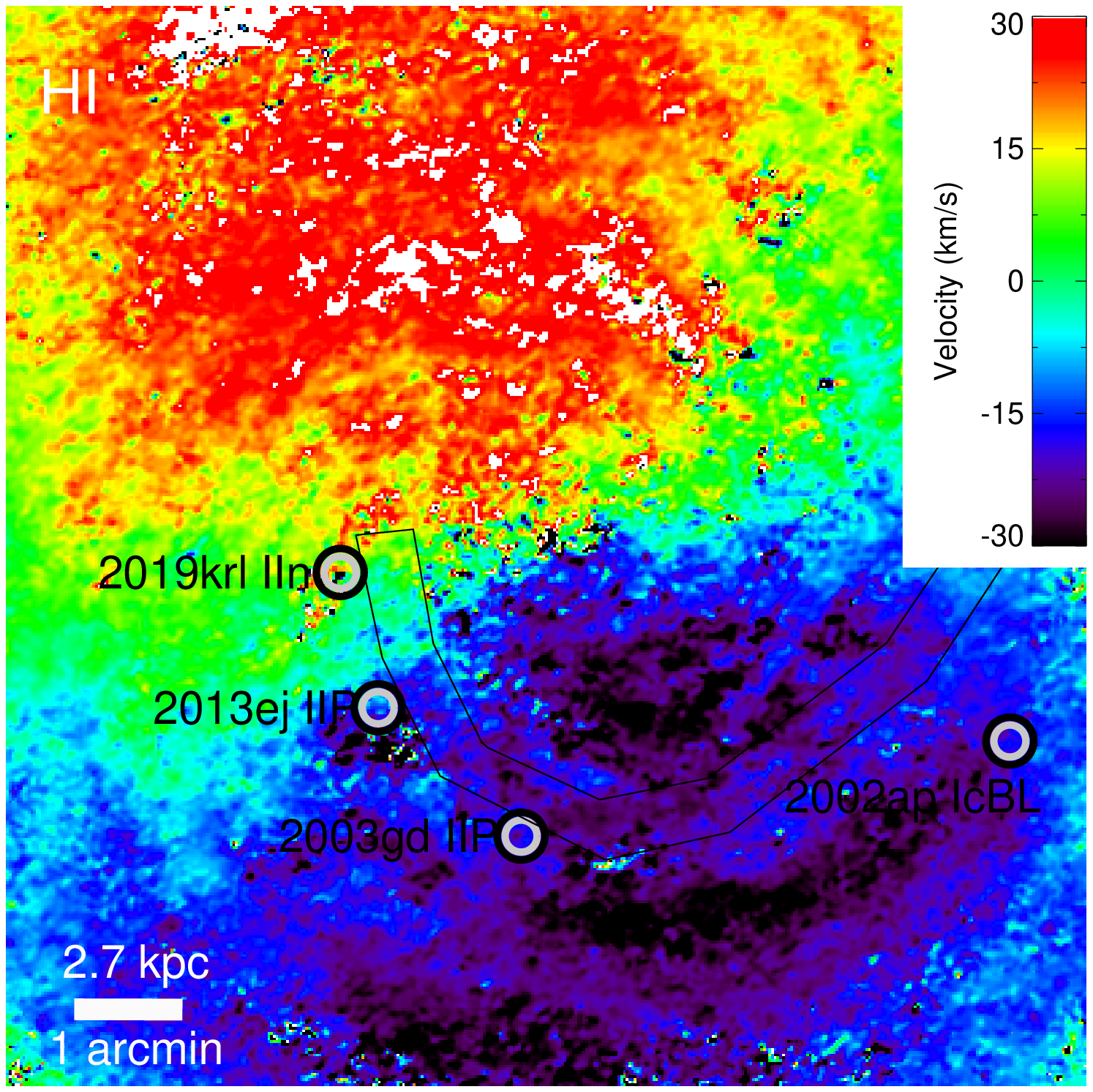} & 
\includegraphics[width=\szerkol,clip]{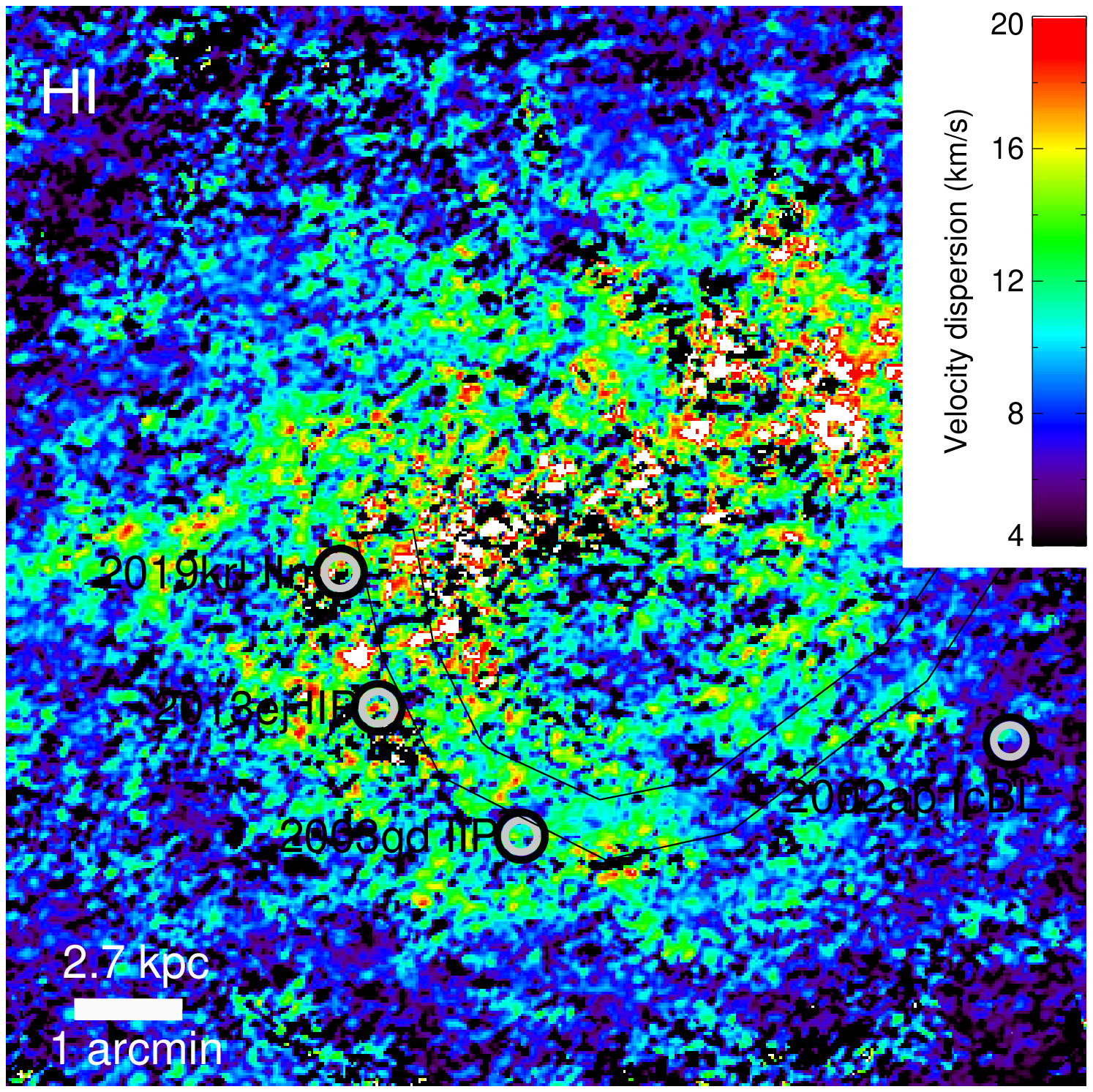} \\
\includegraphics[width=\szerkol,clip]{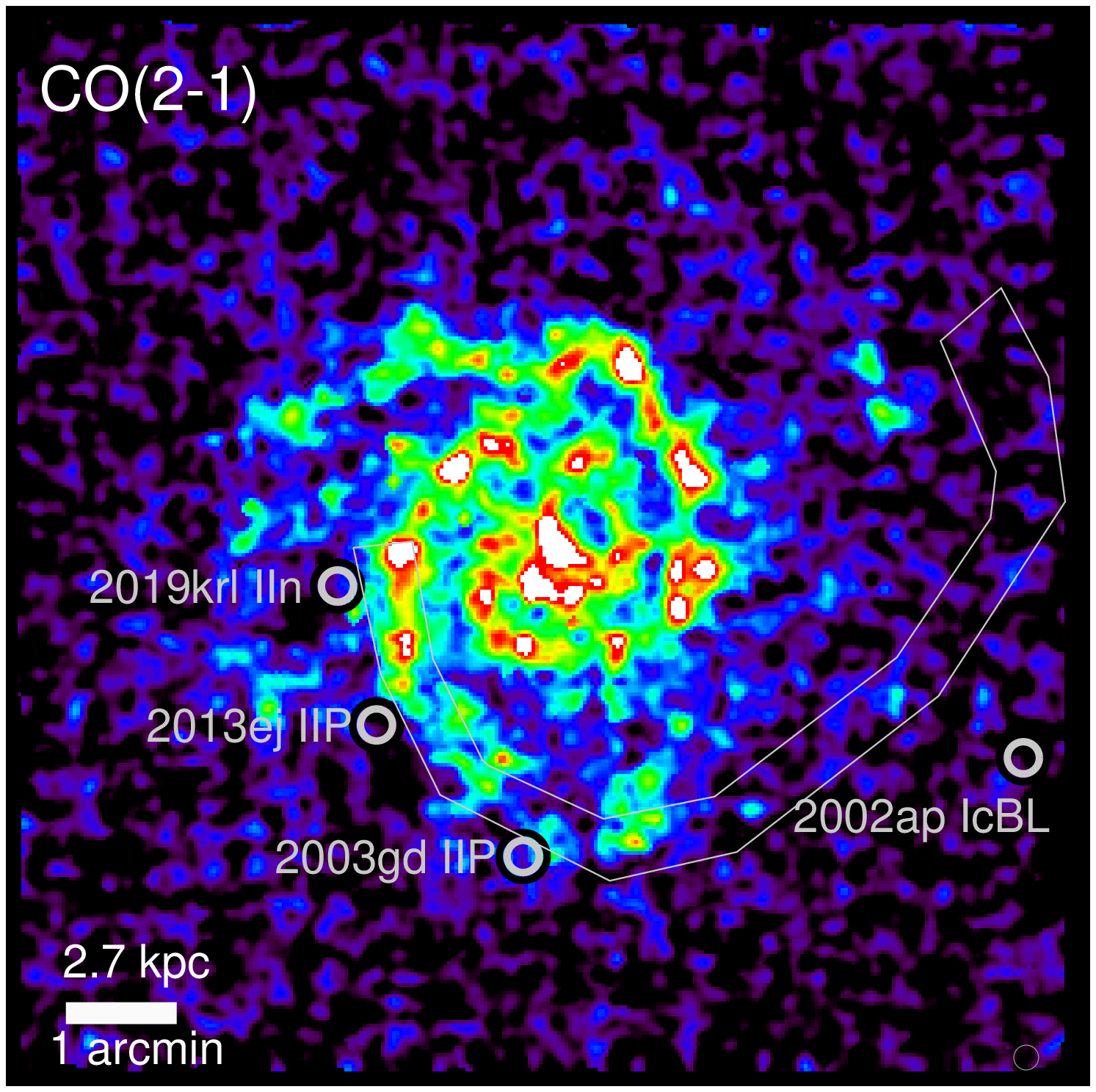} & 
\includegraphics[width=\szerkol,clip]{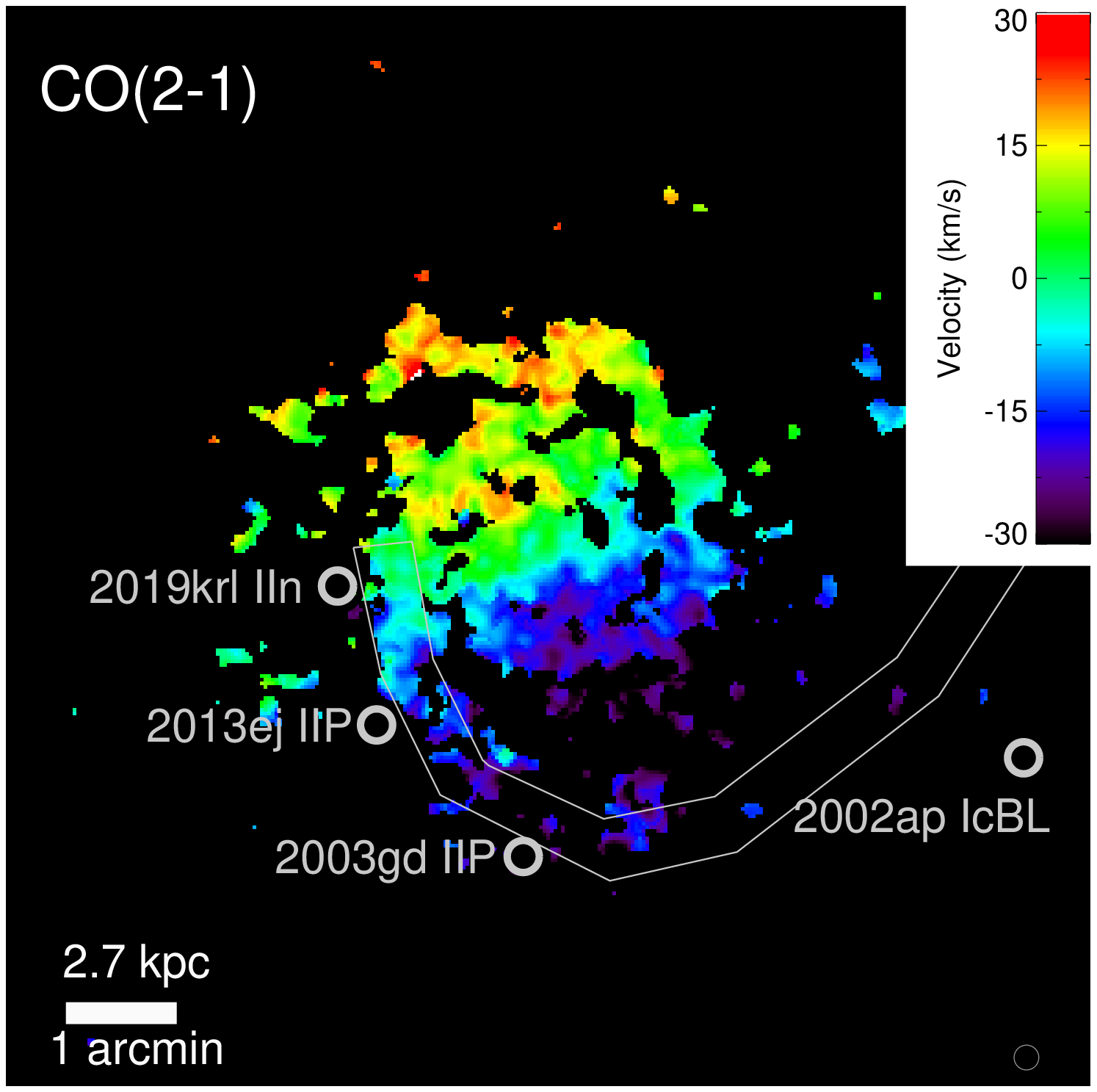} & 
\end{tabular}
\end{center}
\caption{Gas distribution in M74. 
Top and middle: 
{\hi} data with a resolution of 
$11.9\arcsec\times9.3\arcsec$ and
$6.9\arcsec\times5.6\arcsec$, respectively \citep{walter08}. Bottom: CO(2-1) data with a resolution of $13.4\arcsec$ \citep{leroy09}. 
Left: Zeroth moment maps (integrated emission).
Middle:
First moment maps (velocity fields) relative to $z=0.00219$ (656.545\,{\kms}).
Right: Second moment maps (velocity dispersion). The positions of SNe are marked by grey circles. The lines outline the main spiral arm. 
Each panel is 10{\arcmin} per side. North is up and east is to the left.
}
\label{fig:image}
\end{figure*}

\begin{figure*}[t]
\begin{center}
\begin{tabular}{ccc}
\includegraphics[width=\szerkol,clip]{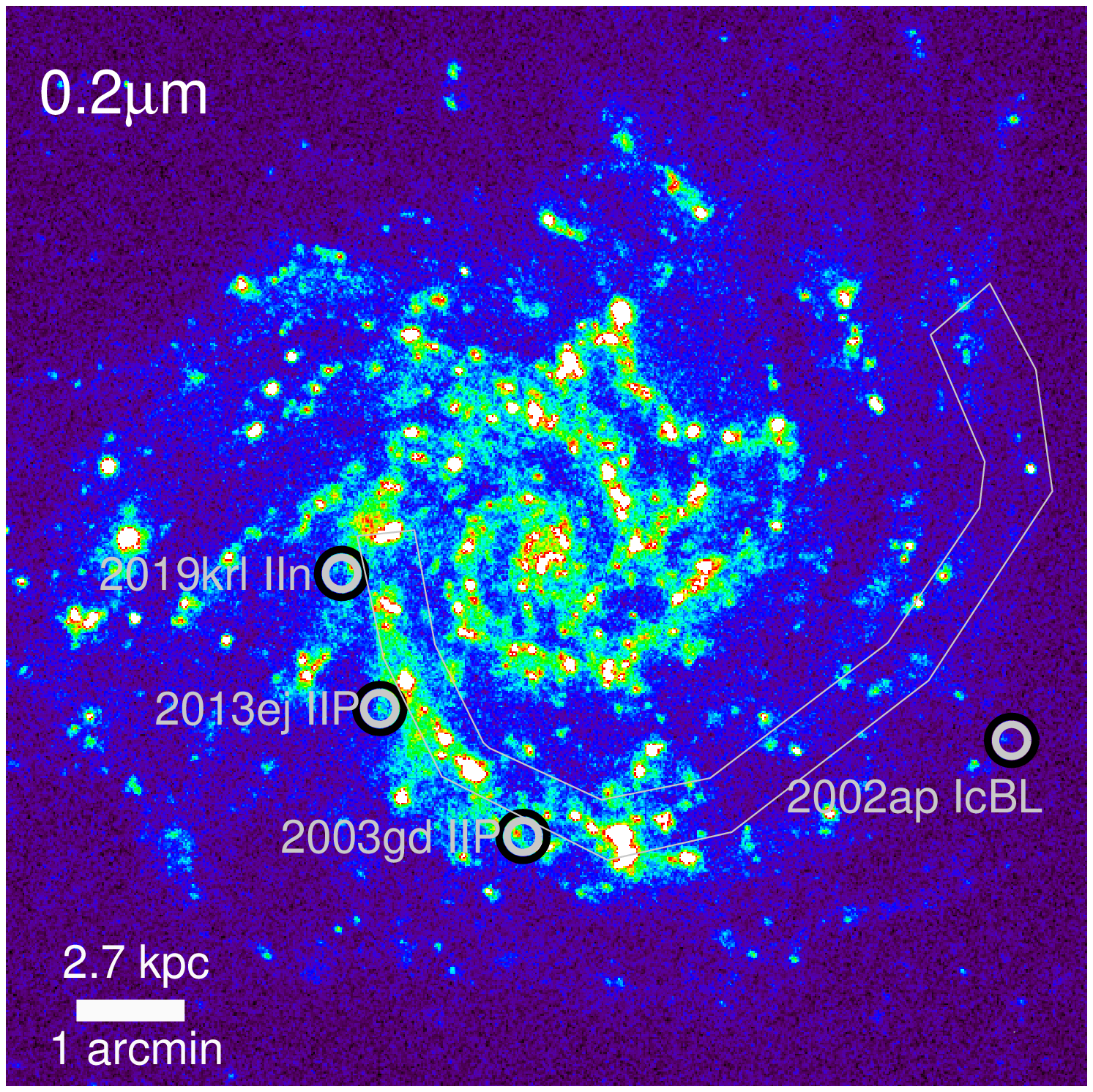} & 
\includegraphics[width=\szerkol,clip]{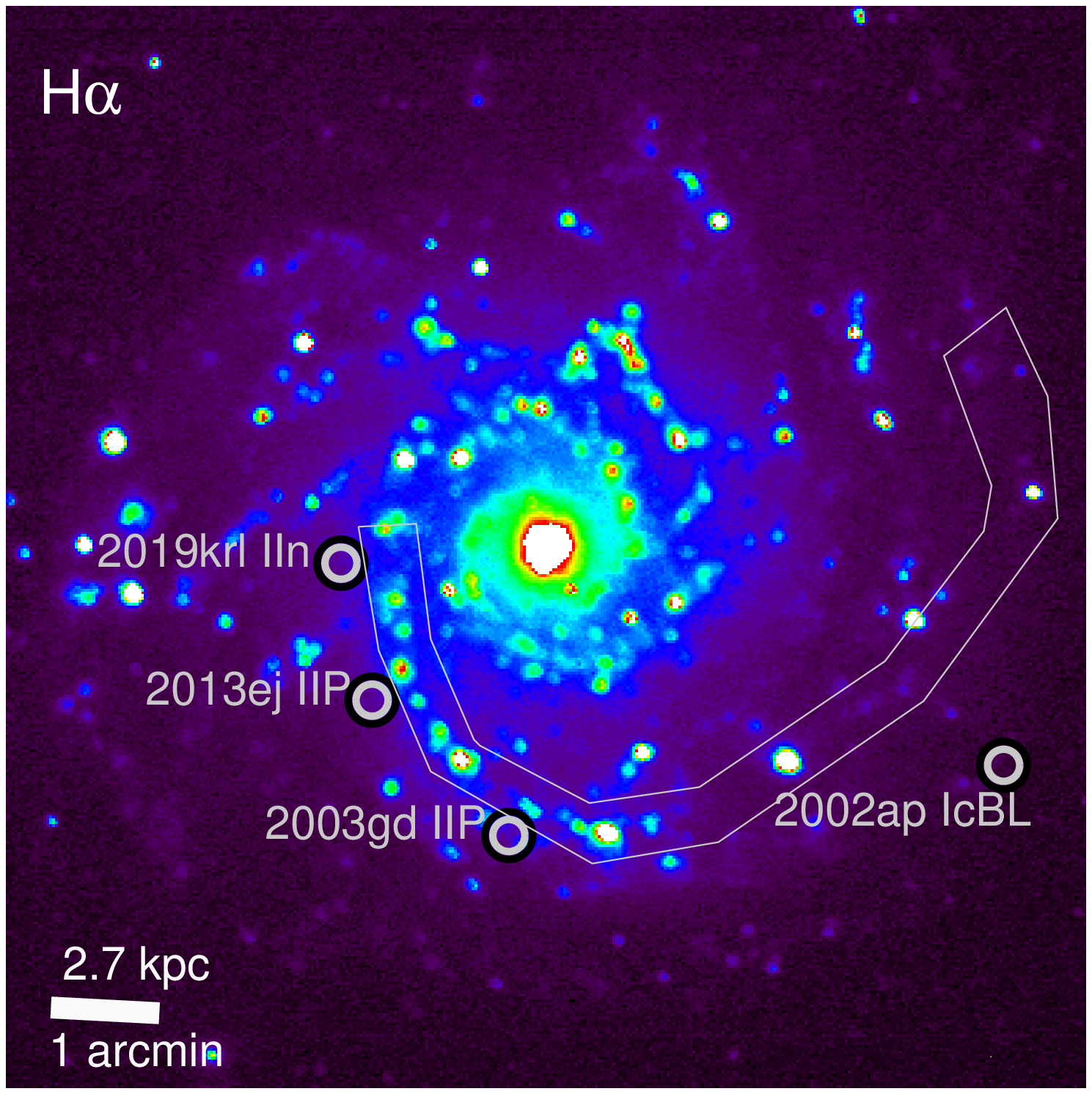} & 
\includegraphics[width=\szerkol,clip]{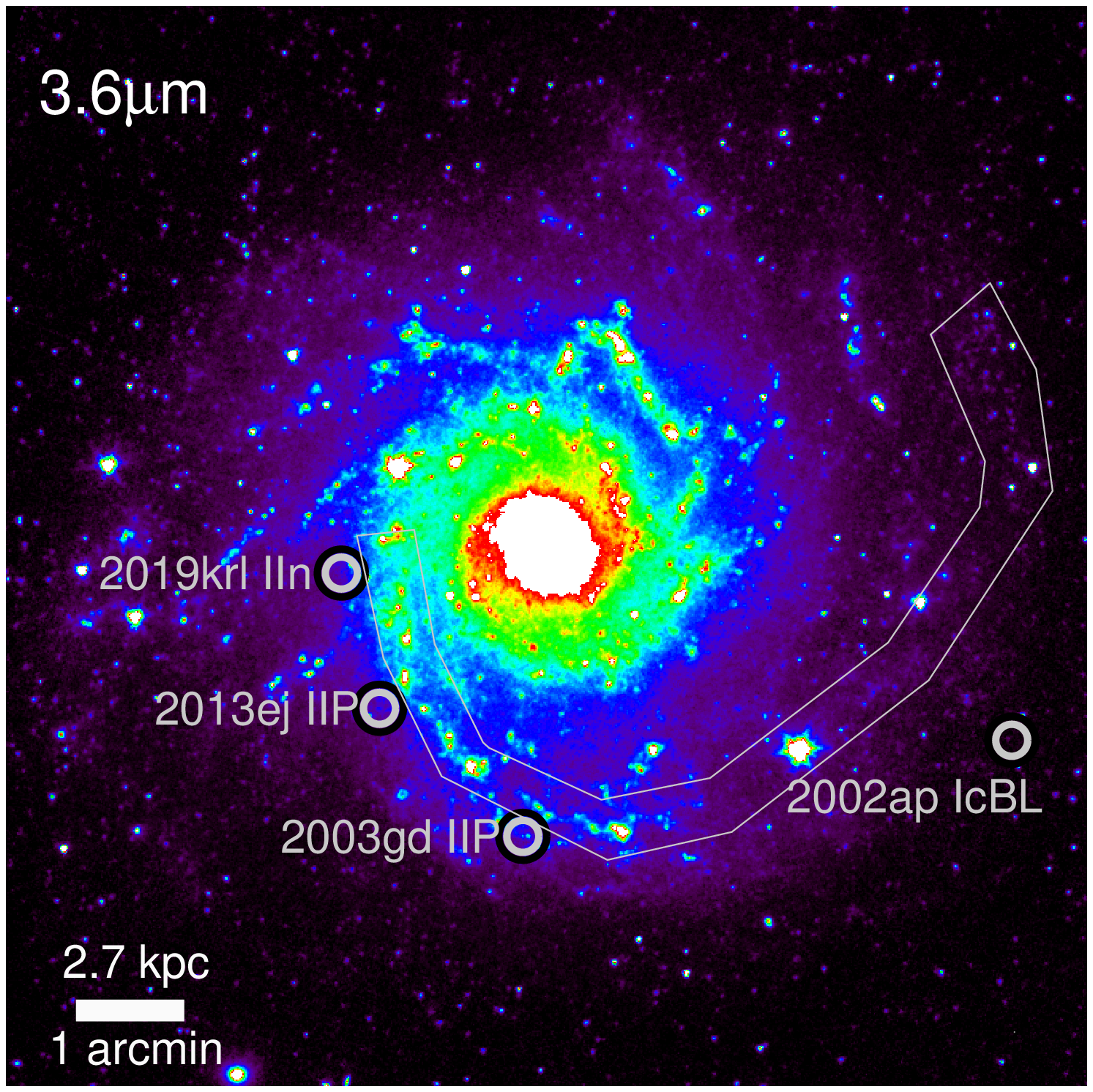} \\
\includegraphics[width=\szerkol,clip]{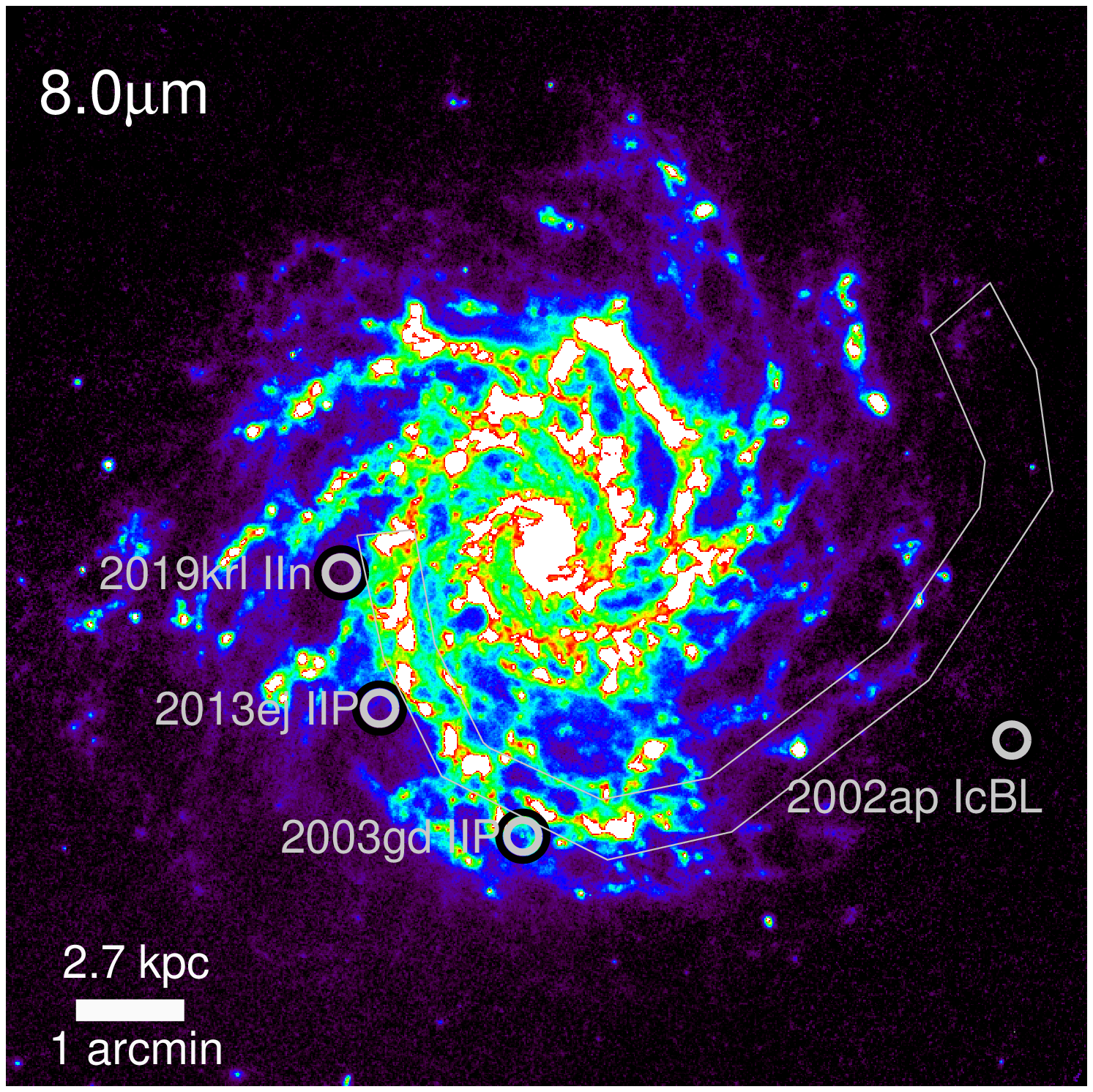} & 
\includegraphics[width=\szerkol,clip]{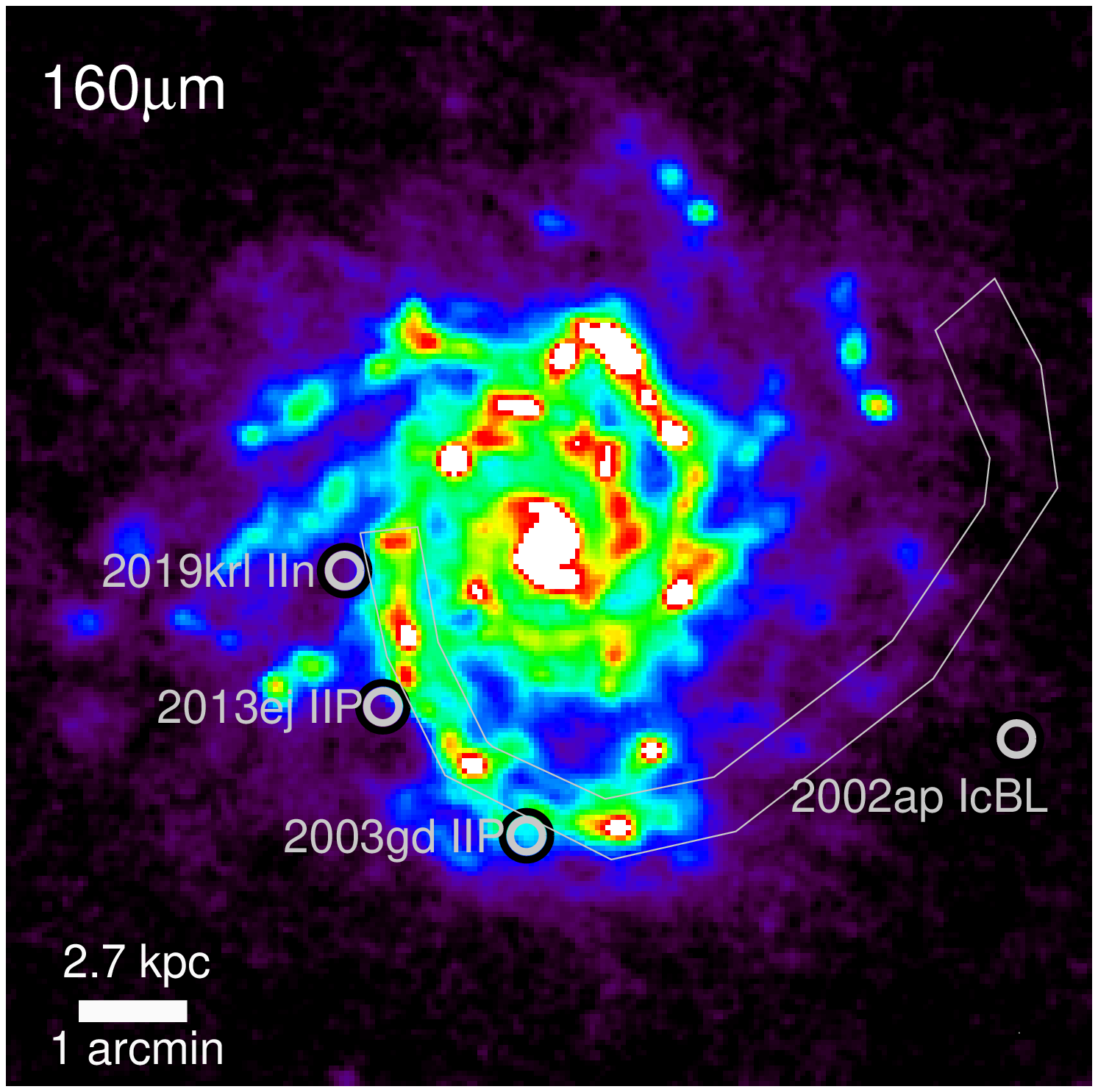} & 
\includegraphics[width=\szerkol,clip]{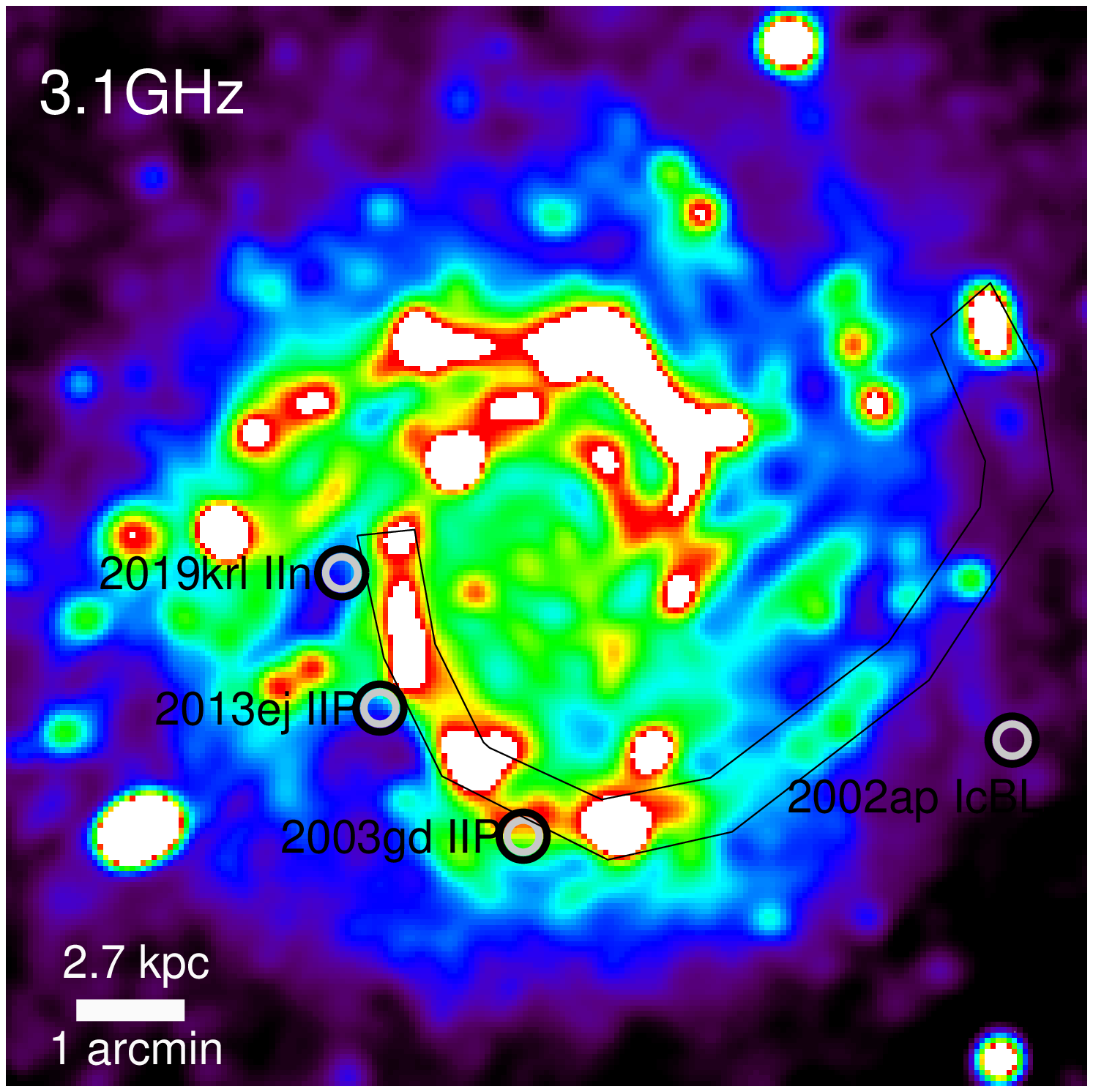} \\
\end{tabular}
\end{center}
\caption{Ultraviolet, H$\alpha$, mid-IR, far-IR, and radio images of M74.
The positions of SNe are marked by grey circles. The lines outline the main spiral arm.
Each panel is 10{\arcmin} per side. North is up and east is to the left.
}
\label{fig:image2}
\end{figure*}

The nature of various types of supernovae (SNe) carry crucial information about stellar evolution.
A subclass of SNe with no hydrogen, helium, or silicon lines in the spectrum (known as type Ic) are believed to be explosions of stars born with very high masses.
Those exhibiting broad emission lines, indicating high velocities of the ejected material (up to a few $10^4\,\kms$), are called `hypernovae' or SNe type Ic-BL (broad lined). Some of these SNe also show relativistic ejecta, for example SN\,1998bw \citep[GRB\,980425][]{galamanature} and 2009bb \citep{soderberg10}. These relativistic features are interpreted as a jet, and indeed some Ic-BL SNe have been associated with gamma-ray bursts (GRBs; \citealt{hjorthsn}).

Observations of atomic and molecular gas (through 21\,cm {\hi} and carbon monoxide [CO] lines, respectively)  in  host galaxies of GRBs and SNe have recently been used to learn about the nature of the explosions themselves, as well as the star formation event during which their progenitors were born.
\citet{michalowski15hi,michalowski16,michalowski18} and \citet{arabsalmani15b,arabsalmani19} showed that GRBs and a relativistic SN type Ic-BL exploded close to the most {\hi}-rich region of their hosts, which was interpreted as being the result of a recent gas accretion or a galaxy merger.
While this conclusion was based on very small samples, if substantiated it would have important consequences for our understanding of the conditions necessary for such explosions, as well as for triggering star formation in general. This motivates us to study the gas properties of another Ic-BL SN (to date, amongst the hosts of Ic-BL SNe, atomic gas was studied in only one case;
\citealt{michalowski18}).

{\sn}  \citep{nakano12circ} 
exploded $\sim4.7\arcmin$ ($\sim12.7$\,kpc)
from the centre of \object{Messier 74} (\object{M74} or \object{NGC\,628}) 
and was classified as a type Ic-BL \citep{mazzali02,kinugasa02,
galyam02,foley03%
}. Its estimated progenitor mass is $20$--$25\,\msun$, lower than other hypernovae, including SN\,1998bw \citep{mazzali02}. {\sn} was also shown to have only modest relativistic ejecta, and hence no detectable jet \citep{berger02}. The progenitor was proposed to be a Wolf-Rayet (WR) star or a massive star in an interacting binary \citep{smartt02,wang03%
}.

M74 has hosted three other known SNe: 2003gd (type IIP; $\sim2.7\arcmin$ or $\sim7.3$\,kpc from the galaxy centre; \citealt{hendry05}), 2013ej (IIP; $\sim2.2\arcmin$ or $\sim5.9$\,kpc;  \citealt{valenti14}), and 2019krl (IIn; $1.9\arcmin$ or $\sim5.2$\,kpc; \citealt{ho19rep,
andrews19atel}). The progenitor of SN\,2003gd was confirmed to be an M-type supergiant with a mass of $\sim8\,\msun$,
by examining the SN position in pre- and post-explosion images, which revealed that this star was missing in the latter
\citep{maund09,vandyk03,smartt04}. In a similar way, the mass of the progenitor of SN\,2013ej was estimated to be $8.0$--$15.5\,\msun$\citep{fraser14,mauerhan17}.
 
The objectives of this paper are to {\it i}) test whether {\sn} and other SNe in M74 were born in concentrations of gas indicating a recent gas accretion, and {\it ii}) investigate what this tells us about the formation of SN progenitors.
We adopt a redshift of M74 of $z=0.00219$ \citep{lu93}, a distance of 9.4\,Mpc, and a corresponding scale of 2.7 kpc arcmin$^{-1}$.
This  assumes a cosmological model with $H_0=70$ km s$^{-1}$ Mpc$^{-1}$,  $\Omega_\Lambda=0.7$, and $\Omega_{\rm m}=0.3$.

\section{Selection and data}
\label{sec:data}

{The supernova \sn} and its host galaxy M74 were selected as part of a larger study of gas in SN hosts (Gotkiewicz \& Micha\l owski, in prep.). We investigated all known SNe up to Aug 2018 with redshifts $z<0.1$ from the Open Supernova Catalog\footnote{{\tt https://sne.space}} \citep{snespace} and searched the NASA/IPAC Extragalactic Database (NED) for {\hi} data for their hosts.  {\sn} was the only SN type Ic-BL that satisfied these criteria.

The archival {\hi} and CO data for M74 are shown in Figs.~\ref{fig:lowres} and \ref{fig:image}, while the rest of the data are shown in Fig.~\ref{fig:image2}.
{\hi} data are from \citet[][a resolution of $72\arcsec\times62\arcsec$]{kamphuis92}  and The {\hi} Nearby Galaxy Survey \citep[THINGS;][a resolution of $11.9\arcsec\times9.3\arcsec$ and
$6.9\arcsec\times5.6\arcsec$]
{walter08}. The CO(2-1) data are from the HERA CO Line Extragalactic Survey \citep[HERACLES;][a resolution of $13.4\arcsec$]{leroy09}. The H$\alpha$ image is from \citet{marcum01}. In addition, we use the following continuum data: 
the {\it Swift} \citep{swift,uvot} 
UVW2 0.2\,{\micron} image \citep{brown14};
{\it Spitzer} \citep{spitzer,irac} 
3.6 and 8.0\,{\micron} images \citep{dale09},
 {\it Herschel}
\citep{herschel,pacs} 
160\,{\micron} image \citep{kingfish}, and
National Science Foundation's (NSF's) Karl G. Jansky Very Large Array (VLA) 3.1\,GHz image \citep{mulcahy17}.

\section{Results}
\label{sec:results}

According to \citet{kamphuis92}, M74 harbours an off-centre asymmetric {\hi} tail, located 
on the south-western outskirts of the galaxy, outside the optical disc
(Fig.~\ref{fig:lowres}). The feature is detected over 20{\arcmin} (55\,kpc) and contains 
$\log(\mhi/\msun)=8.95$, or 7.5\% of the total atomic gas of M74. {\sn} is located 
where this feature connects with the symmetric disc of M74.
The velocity pattern of the feature is irregular 
(Fig.~\ref{fig:lowres}).
This gas does not follow the overall rotation of the gas disc, as evidenced by both negative and positive velocity residuals from the disc models at this location presented in Figs.~8 and 9 of \citet{kamphuis92}. 

The higher resolution THINGS data are not sensitive to such large scales, but allow detailed investigation of the local environments of SNe.
At this resolution, the position of {\sn} is not associated with any strong concentration of atomic or molecular gas (Fig.~\ref{fig:image}).
{\sn} is located $\sim80\arcsec$ (3.6\,kpc) south-west of the main spiral arm running from the south to the west of the galaxy (marked as a curved region on Figs.~\ref{fig:image} and \ref{fig:image2}, traced clearly on all images up to the southernmost point, and by the {\hi} and 3.1\,GHz images to the west) 
and $\sim20\arcsec$ (0.9\,kpc) from 
a bright {\hi} knot to the north. The main spiral arm in the south is also visible in the {\hi} velocity map in which 
the integrated mean velocities show a larger deviation from the systemic velocity in the interarm regions.

Moreover, {\sn} exploded away from regions of significant star formation activity and very little emission is present at its position at any wavelength (Fig.~\ref{fig:image2}). There is, however, a faint star-forming region visible in the UV $\sim4\arcsec$ (180\,pc) away from the SN position (this is not the object $10\arcsec$ away mentioned by \citealt{crowther13}).
{\sn} is also outside the CO disc.

The other three type II SNe in M74 exploded along the most prominent spiral arm (running from the east to south of M74; curved regions on Figs.~\ref{fig:image} and \ref{fig:image2}), but are displaced from the arm towards the outside 
by $\sim25\arcsec$ ($\sim1$\,kpc). This is especially evident in the H$\alpha$, $3.6\,\micron,$ and CO images. SN\,2013ej and 2019krl exploded in interarm regions with very little {\hi}, 8, 160\,{\micron,} and 3.1\,GHz emission.  All three type II SNe exploded in regions with undetectable CO emission.

We have investigated the large-scale environment of M74. Within 150\,kpc (55{\arcmin}) and $\pm500\,\kms$ from M74 (velocity of $627\,\kms$), NED lists three galaxies:
UGC\,1171, 
UGC\,1176, 
both to the east,
and the much fainter SDSS J013800.30+145858.1 
to the south. 
All of them are detected in {\hi} by the Arecibo Legacy Fast ALFA Survey (ALFALFA; \citealt{haynes18}). In Table~\ref{tab:other} we list their properties.

\begin{figure}
\begin{center}
\includegraphics[width=0.5\textwidth,clip]{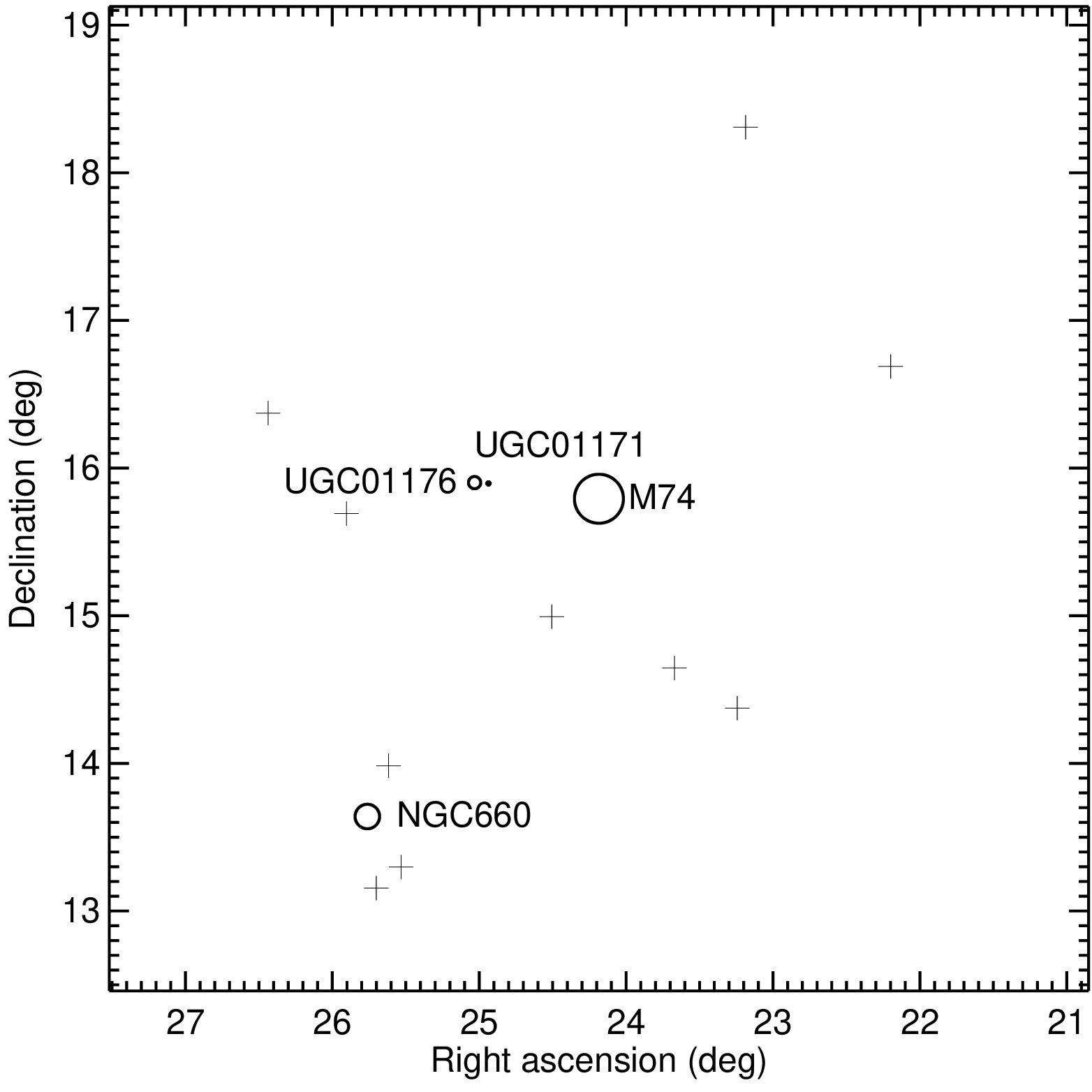} 
\end{center}
\caption{Large-scale environment around M74 within $\pm500\,\kms$. The width of the panel is 400{\arcmin} ($\sim1$\,Mpc). Circles show the sizes of larger galaxies: 20{\arcmin} for M74, 10{\arcmin} for NGC\,660, 5{\arcmin} for UGC\,1176, and 1.4{\arcmin} for UGC\,1171 (a small circle next to UGC\,1176). Crosses show the positions of additional galaxies detected at the {\hi} line by ALFALFA \citep{haynes18}. 
}
\label{fig:comp}
\end{figure}

In the ALFALFA catalogue within 200{\arcsec} (540\,kpc) of M74 there are in total 13 galaxies. All but one have atomic gas masses $7<\log(\mhi/\msun)<9$. Only NGC\,660, 426\,kpc away, with $\log(\mhi/\msun)=9.59$ has a mass comparable to that of M74.  
The positions of these galaxies are shown on Fig.~\ref{fig:comp}.

\begin{table*}
\caption{Properties of galaxies in the vicinity of M74 from the ALFALFA survey  \citep{haynes18}.}
    \centering
    \begin{tabular}{lrccrccccc}
    \hline\hline
    Galaxy & ID & RA & Dec & \multicolumn{2}{c}{Dist$_{\rm M74}$} & $z$ & $V_{\rm helio}$ & $f_{HI}$ & $\log(\mhi)$ \\ 
     & & (deg) & (deg) & ($'$) & (kpc) & & ($\kms$) & (Jy\,$\kms$) & ($\msun$) \\
     \hline
    M74 & 1149 & 24.18500 & 15.79222 & $\cdots$ & $\cdots$ & 0.002190 & 657 & $424.30\pm0.18$ & $9.73\pm0.1\phantom{0}$ \\
    UGC\,1171 & 1171 & 24.93708 & 15.89611 & 44.6 & 120 & 0.002463 & 738 & $\phantom{10}2.07\pm0.05$ & $7.42\pm0.09$ \\
    UGC\,1176 & 1176 &  25.03167 & 15.90167 &  50.6 & 137 & 0.002103 & 630 & $\phantom{1}31.22\pm0.06$ & $8.78\pm0.1\phantom{0}$ \\
    SDSSJ0138
    & 112503 & 24.50583 & 14.99333 & 51.6 & 139 & 0.002478 & 743 & $\phantom{10}0.56\pm0.05$ & $7.14\pm0.2\phantom{0}$ \\
    NGC\,660 & 1201 & 25.76167 & 13.64000 &  157.9 & 426 & 0.002830 & 848 & $148.39\pm0.14$ & $9.59\pm0.34$ \\
    \hline
    \end{tabular}
    \label{tab:other}
    \tablefoot{The columns show the galaxy name (that of SDSSJ013800.30+145858.1 has been abbreviated), ALFALFA ID, position, projected distance to M74, redshift, heliocentric velocity, {\hi} flux, and the atomic gas mass.}
\end{table*}

\section{Discussion}
\label{sec:discussion}

The estimated mass of the progenitor of  {\sn} implies that it was formed 3.2--5.6\,Myr before the explosion, whereas the progenitors of 
SN\,2003gd and 2013ej
were formed 10--55\,Myr before the explosions 
(assuming a main-sequence lifetime of $10^{10}\,\mbox{yr} \times [M/\msun\mbox{$]$} ^{-2.5}$; \citealt{kippenhahn90}).
This lifetime for {\sn} agrees with the single-progenitor estimate of \citet[][5\,Myr]{zapartas17}, but is lower than the binary-progenitor estimate (20\,Myr). Similarly, \citet{maund18} obtained an age of 15\,Myr based on the analysis of stars in the vicinity of {\sn}.

\subsection{Type Ic-BL {\sn}}

{The supernova \sn} is the fourth known explosion of a presumably massive star located close to a concentration of atomic gas. Similarly to {\sn},  GRB\,060505 and SN\,2009bb were both located close to {\hi} extensions, whereas GRB\,980425 was located close to the most significant {\hi} concentration \citep{michalowski15hi,michalowski18,arabsalmani15b,arabsalmani19}.

We can assess the statistical significance of the associations of the {\hi} off-centre concentrations with the GRB and SN positions by investigating the probability of four GRBs and SNe exploding by chance in the quadrants of their hosts at which these {\hi} concentrations are located. For a given explosion this probability is $0.25$, so for four out of four analysed cases this is $(0.25)^4\sim0.004$. This corresponds to $\sim3\sigma$, so the associations are unlikely to be random, but a larger sample is needed to confirm this result.

The case of {\sn} adds to 
the hypothesis put forward in \citet{michalowski15hi} that the progenitors of these explosions are born when atomic gas is accreted from the intergalactic medium. Indeed, \citet{kamphuis92} concluded that the south-western {\hi} extension in M74 is a result of an accretion event 
because it has not settled yet, and is inconsistent with the rotation of the gas disc. This extension may be the gas flowing in and feeding star formation directly, or distorting gas on the outskirts of the optical disc of the galaxy leading to star formation at the position of {\sn}.

The asymmetric tail in M74 has a mass of $\log(\mhi/\msun)=8.95$. This is comparable to the mass of the most massive galaxy within 150\,kpc  (UGC\,1176) and only a factor of four less than the atomic gas mass of NGC\,660.
The sum of atomic masses of all galaxies within 200{\arcmin} (540\,kpc) excluding NGC\,660 is $\log(\mhi/\msun)=9.36$.
Hence, the tail in M74 might have come from UGC\,1176 if it was significantly distorted by the interaction and lost half of its gas, or from NGC\,660.
It could also be a remnant of a galaxy similar to UGC\,1176 that has been accreted entirely (as postulated by \citealt{kamphuis92}), or a result of the accretion of intragroup medium.

In principle this tail could be a remnant of a  tidal feature created by interaction with these galaxies, but we found this interpretation unlikely. First, the tail does not resemble recent tidal features, whereas an older feature would  wind almost symmetrically around the galaxy. Second, simulations shows that tidal tails are created on both sides of interacting galaxies \citep{hopkins06b,hayward12,hayward14,pettitt16,oh16}, whereas M74 does not have such feature on the other side.
We also note that the asymmetric nature of M74, with the southern arm being stronger than the northern one (Fig.~\ref{fig:image}), could be a result of interaction with the UGC\,1176/1171 pair.

The only possible counter-example of a potential  explosion of a massive star without an associated gas concentration is the enigmatic transient AT\,2018cow, whose host galaxy does not show such off-centre asymmetric {\hi} features  \citep{michalowski19} and possibly only a gas ring \citep{roychowdhury19}. 
Such an {\hi} ring would also be apparent for M74 if it was further away so the sensitivity and  resolution were poorer, because the central part is devoid of atomic gas, likely due to conversion to the molecular phase (Fig.~\ref{fig:image}).
To demonstrate this we smoothed the {\hi} VLA image with a Gaussian with a full width half maximum of 100{\arcsec} (4.5\,kpc;   Fig.~\ref{fig:smooth} in the appendix). At this resolution the spiral structure of M74 resembles an irregular ring, similar to that detected for the AT\,2018cow host (for which the resolution was around 2\,kpc).
However, the nature of AT\,2018cow is not clear, so it may not be connected with the explosion of a massive star
(\citealt{prentice18,liu18,kuin19,perley19,soker19,lyutikov19,bietenholz20}, but see \citealt{prentice18,riverasandoval18,margutti19,fox19,huang19}).

It is unlikely that the lack of molecular gas or star formation at the position of {\sn} is due to the progenitor being kicked out of a star-forming region. The velocities of runaway stars are up to 200\,{\kms} \citep{blaauw93,%
hoogerwerf01,%
eldridge11%
}, which corresponds to 1\,kpc per 5\,Myr. This is only $\sim20\arcsec$ at the distance of M74, so not sufficient to move the birth place of the {\sn} progenitor to any place of significant star formation or CO concentration.
This is true even if the lifetime of the progenitor is three to four times longer (15--20\,My; \citealt{zapartas17}, \citealt{maund18}).
CO deficiency at  GRB positions was also claimed by \citet{hatsukade14}, \citet{stanway15}, and \citet{michalowski16}, but \citet{perley17}, \citet{michalowski18co}, and \citet{arabsalmani18} suggest an alternative. If the lack of molecular gas is confirmed for a larger sample of type Ic-BL SNe, this would support the hypothesis of {\hi}-fuelled star formation giving rise to the birth of their progenitors \citep{michalowski15hi}.


It is unlikely that the {\sn} progenitor moved to its explosion position due to a random kick.
Assuming a lifetime of 5\,Myr, the {\sn} progenitor could not be born in the main arm, as the required velocity is 700\,{\kms} to cross 3.6\,kpc. Even the closest bright {\hi} knot 
to the north
is likely too far to be the birthplace, as this would require a velocity of 175\,{\kms}. 
Such  velocity kicks are at the high end for runaway stars \citep{hoogerwerf01}. For the longer lifetime estimates of 15--20\,Myr, the required velocities from the spiral arm would be 235--175\,\kms, so still too high for the {\sn} progenitor to have been born there. However, in such a case it is feasible that it was born in the closest bright {\hi} knot to the north,
as this would require velocities of 60--40\,\kms.

\subsection{Type II SN\,2003gd, 2013ej, and 2019krl}
      
All type II SNe in M74 
are not located close to the {\hi} extension, so are unlikely connected to gas accretion.
They are located at the outside edge of a spiral arm. This can be explained by either of two scenarios: 
by a gas density build-up and shock scenario at the edge of the arm, or by SN progenitors moving away from the arm during their lifetimes.

The first possibility is that the SN progenitors are born when gas is piling up and shocked at the edge of the arm when gas clouds are being swept up by the arm, as explained by the spiral density wave theory (\citealt{shu16}). This is similar to the hypothesis presented in \citet{michalowski14} that GRB progenitors are preferentially born in high-density gas.
We note that the amount of gas piling up at the edge of the arm giving rise to the birth of SN progenitors cannot be large, because the concentration is not detected with {\hi} or CO observations. 

The SNe in M74 are on the outside of the spiral arm, so this scenario is only valid if they are outside the corotation radius (where the orbital velocity is equal to the spiral pattern speed), so the arm is catching up with gas that is moving slower \citep{shu16}. 
\citet{aramyan16} found that core-collapse SNe are indeed shifted towards the outside edge of spiral arms as long as they are  outside the corotation radius.
Unfortunately the accuracy of the estimate of the corotation radius for M74 is not sufficient to test this. The corotation radius given by \citet{scarano13}\footnote{$4.6\pm1.2$\,kpc and their adopted scale of 1.95 kpc arcmin$^{-1}$.} is $(2.4\pm0.6)\arcmin$, corresponding to $(6.4\pm1.7)\,$kpc with our adopted distance, so it cannot be established whether type II SNe are inside or outside this radius. 
\citet{karapetyan18} quoted a slightly lower (but consistent within errors) value of the corotation radius\footnote{The ratio of the corotation and isophotal  ($R_{25}=5.52\arcmin$) radii of  $0.34\pm0.09$.} of $(1.9\pm0.5)\arcmin$, concluding that SNe\,2003gd and 2013ej are  indeed outside the corotation radius, supporting the spiral density wave scenario for their formation.
This scenario cannot explain the birth of the {\sn} progenitor, because gas should not pile up so far away (3.6\,kpc; Fig.~\ref{fig:image}) from the spiral arm.


However, the region of increased gas density and shocks, predicted by the spiral density wave theory, does manifest itself with increased star formation, and therefore more intense UV and H$\alpha$ emission. These main sites of star formation are where SN progenitors should be born, not 1\,kpc away. This scenario does not explain this discrepancy.

The second possibility is that SN progenitors are in fact born in the spiral arm, but, due to their orbital motions, move away before they explode. 
Inside the corotation radius stars (and gas) are moving faster than the spiral pattern, so are constantly drifting towards the outside of the arm. 
This means that SNe with progenitors with long enough lifetimes should be happening preferentially outside the arm. On the other hand, H$\alpha$ emission is dominated by stars born very recently (i.e. after the SN progenitors), which have therefore not yet moved away from the arm.

The lifetimes of type II progenitors imply that they would need to have (reasonable) velocities of $18$--$100\,\kms$ with respect to the arm to cross 1\,kpc from the arm to their current position.
We note that this velocity is not due to any random-direction kick, but is the orbital velocity minus the spiral pattern speed.




The orbital migration scenario requires that type II SNe progenitors are inside the corotation radius, so their orbital motion is faster than the spiral pattern and so they can leave it on the outside edge \citep[e.g.][]{aramyan16}.
This effect is also visible in M51 where younger stellar clusters have a distribution that has the strongest correlation with the distribution of star formation, and they are shifted towards the outside of the arm inside the corotation radius \citep{scheepmaker09}.

This scenario cannot explain the position of {\sn}, because it is located securely outside the corotation radius, so cannot leave the spiral arm at the outside edge. Instead, as we discuss above, it may
be connected with the gas accretion visible in the {\hi} map of \citet{kamphuis92}.

The position of SNe away from the main sites of star formation (spiral arms) is not unique to M74. We investigated galaxies with four or more core-collapse SNe and with well-separated spiral structures from the list of \citet{thone09}. NGC\,6946 and 4303 hosted nine and six type II SNe, respectively, and only one (1948B) and two (1999gn and 2006ov) exploded in spiral arms; the rest exploded in interarm regions or outside the detectable stellar disk (Figs.~9 and 11 of \citealt{anderson13}). 
This is consistent with the spiral density wave theory and the migration scenario described above.
Indeed, larger samples of type II SNe show that they do not concentrate in the brightest regions of their hosts \citep{fruchter06,anderson08,
leloudas11}.

\section{Conclusions}
\label{sec:conclusion}

We have used archival {\hi} and CO data for M74 (not previously investigated in the context of SN positions), together with H$\alpha$ and continuum images.
{\sn} is located  at  the end of  an  off-centre asymmetric 55\,kpc-long {\hi} extension containing 7.5\% of the total atomic gas of M74. 
It is the fourth known explosion of a presumably massive star that is located close to the concentration of atomic gas (after GRB\,980425, 060505, and SN\,2009bb). 
It is unlikely that all these associations are random (at a $3\sigma$ significance), so
the case of {\sn} adds to
 the evidence that the birth of the progenitors of type Ic-BL SNe and GRBs is connected with the accretion of atomic gas from the intergalactic medium.
The {\hi} extension could come from tidally disrupted companions of M74, or be a remnant of a galaxy or a gas cloud in the intragroup medium accreted entirely. 

The type II SNe in M74 do not seem to be related to gas accretion.
The fact that 
they
are located at the outside edge of a spiral arm suggests either that their progenitors are born when gas is piling up there, reaching high density, or that SN progenitors move away from the arm during their lifetimes, due to their orbital motions.
This is also similar for NGC\,6946 and 4303, with eight out of nine and four out of six type II SNe, respectively, located in interarm regions or outside the detectable stellar discs, not in the spiral arms.

\begin{acknowledgements}

We wish to thank the referee for careful and important suggestions, Joanna Baradziej and Phillip Hopkins  for discussion and comments, and Frank Briggs for permission to use the figure from \citet{kamphuis92}.

M.J.M.~acknowledges the support of 
the National Science Centre, Poland through the SONATA BIS grant 2018/30/E/ST9/00208, and of the Polish-U.S. Fulbright Commission.
J.H.~was supported by a VILLUM FONDEN Investigator grant (project number 16599).
P.K.~is partially supported by the BMBF project 05A17PC2 for D-MeerKAT.
The National Radio Astronomy Observatory is a facility of the National Science Foundation operated under cooperative agreement by Associated Universities, Inc. 
This work made use of HERACLES, `The HERA CO-Line Extragalactic Survey'. 
This research has made use of data obtained from the High Energy Astrophysics Science Archive Research Center (HEASARC), provided by NASA's Goddard Space Flight Center. 
This work is based on observations made with the Spitzer Space Telescope, which is operated by the Jet Propulsion Laboratory, California Institute of Technology under a contract with NASA. 
{\it Herschel} is an ESA space observatory with science instruments provided by European-led Principal Investigator consortia and with important participation from NASA. 
This research has made use of 
the Open Supernova Catalog (\url{https://sne.space});
NASA/IPAC Extragalactic Database (NED), which is operated by the Jet Propulsion Laboratory, California Institute of Technology, under contract with the National Aeronautics and Space Administration;
SAOImage DS9, developed by the Smithsonian Astrophysical Observatory \citep{ds9};
Edward Wright cosmology calculator \citep{wrightcalc};
the WebPlotDigitizer of Ankit Rohatgi ({\tt arohatgi.info/WebPlotDigitizer});
and NASA's Astrophysics Data System Bibliographic Services.
\end{acknowledgements}



\begin{thebibliography}{85}
\expandafter\ifx\csname natexlab\endcsname\relax\def\natexlab#1{#1}\fi
\expandafter\ifx\csname url\endcsname\relax
  \def\url#1{{\tt #1}}\fi
\expandafter\ifx\csname urlprefix\endcsname\relax\def\urlprefix{URL }\fi

\bibitem[{{Anderson} \& {James}(2008)}]{anderson08}
{Anderson} J.P., {James} P.A., 2008, \mnras, 390, 1527

\bibitem[{{Anderson} \& {Soto}(2013)}]{anderson13}
{Anderson} J.P., {Soto} M., 2013, \aap, 550, A69

\bibitem[{{Andrews} et~al.(2019){Andrews}, {Sand}, {Smith}
  et~al.}]{andrews19atel}
{Andrews} J., {Sand} D., {Smith} N., {Moe} M., {Lundquist} M., {Kattner} S.,
  2019, The Astronomer's Telegram, 12913

\bibitem[{{Arabsalmani} et~al.(2015){Arabsalmani}, {Roychowdhury}, {Zwaan},
  {Kanekar}, \& {Micha{\l}owski}}]{arabsalmani15b}
{Arabsalmani} M., {Roychowdhury} S., {Zwaan} M.A., {Kanekar} N.,
  {Micha{\l}owski} M.J., 2015, \mnras, 454, L51

\bibitem[{{Arabsalmani} et~al.(2018){Arabsalmani}, {Le Floc'h}, {Dannerbauer}
  et~al.}]{arabsalmani18}
{Arabsalmani} M., et~al., 2018, \mnras, 476, 2332

\bibitem[{{Arabsalmani} et~al.(2019){Arabsalmani}, {Roychowdhury},
  {Starkenburg} et~al.}]{arabsalmani19}
{Arabsalmani} M., et~al., 2019, \mnras, 485, 5411

\bibitem[{{Aramyan} et~al.(2016){Aramyan}, {Hakobyan}, {Petrosian}
  et~al.}]{aramyan16}
{Aramyan} L.S., et~al., 2016, \mnras, 459, 3130

\bibitem[{{Berger} et~al.(2002){Berger}, {Kulkarni}, \& {Chevalier}}]{berger02}
{Berger} E., {Kulkarni} S.R., {Chevalier} R.A., 2002, \apjl, 577, L5

\bibitem[{{Bietenholz} et~al.(2020){Bietenholz}, {Margutti}, {Coppejans}
  et~al.}]{bietenholz20}
{Bietenholz} M.F., et~al., 2020, \mnras, 491, 4735

\bibitem[{{Blaauw}(1993)}]{blaauw93}
{Blaauw} A., 1993, In: {Cassinelli} J.P., {Churchwell} E.B. (eds.) Massive
  Stars: Their Lives in the Interstellar Medium, vol.~35 of Astronomical
  Society of the Pacific Conference Series, 207

\bibitem[{{Brown} et~al.(2014){Brown}, {Moustakas}, {Smith} et~al.}]{brown14}
{Brown} M.J.I., et~al., 2014, \apjs, 212, 18

\bibitem[{{Crowther}(2013)}]{crowther13}
{Crowther} P.A., 2013, \mnras, 428, 1927

\bibitem[{{Dale} et~al.(2009){Dale}, {Cohen}, {Johnson} et~al.}]{dale09}
{Dale} D.A., et~al., 2009, \apj, 703, 517

\bibitem[{{Eldridge} et~al.(2011){Eldridge}, {Langer}, \& {Tout}}]{eldridge11}
{Eldridge} J.J., {Langer} N., {Tout} C.A., 2011, \mnras, 414, 3501

\bibitem[{{Fazio} et~al.(2004){Fazio}, {Hora}, {Allen} et~al.}]{irac}
{Fazio} G.G., et~al., 2004, \apjs, 154, 10

\bibitem[{{Foley} et~al.(2003){Foley}, {Papenkova}, {Swift} et~al.}]{foley03}
{Foley} R.J., et~al., 2003, \pasp, 115, 1220

\bibitem[{{Fox} \& {Smith}(2019)}]{fox19}
{Fox} O.D., {Smith} N., 2019, \mnras, 488, 3772

\bibitem[{{Fraser} et~al.(2014){Fraser}, {Maund}, {Smartt} et~al.}]{fraser14}
{Fraser} M., et~al., 2014, \mnras, 439, L56

\bibitem[{{Fruchter} et~al.(2006){Fruchter}, {Levan}, {Strolger}
  et~al.}]{fruchter06}
{Fruchter} A.S., et~al., 2006, \nat, 441, 463

\bibitem[{{Gal-Yam} et~al.(2002){Gal-Yam}, {Ofek}, \& {Shemmer}}]{galyam02}
{Gal-Yam} A., {Ofek} E.O., {Shemmer} O., 2002, \mnras, 332, L73

\bibitem[{{Galama} et~al.(1998){Galama}, {Vreeswijk}, {van Paradijs}
  et~al.}]{galamanature}
{Galama} T.J., et~al., 1998, \nat, 395, 670

\bibitem[{{Gehrels} et~al.(2004){Gehrels}, {Chincarini}, {Giommi}
  et~al.}]{swift}
{Gehrels} N., et~al., 2004, \apj, 611, 1005

\bibitem[{{Guillochon} et~al.(2017){Guillochon}, {Parrent}, {Kelley}, \&
  {Margutti}}]{snespace}
{Guillochon} J., {Parrent} J., {Kelley} L.Z., {Margutti} R., 2017, \apj, 835,
  64

\bibitem[{{Hatsukade} et~al.(2014){Hatsukade}, {Ohta}, {Endo}
  et~al.}]{hatsukade14}
{Hatsukade} B., {Ohta} K., {Endo} A., {Nakanishi} K., {Tamura} Y., {Hashimoto}
  T., {Kohno} K., 2014, \nat, 510, 247

\bibitem[{{Haynes} et~al.(2018){Haynes}, {Giovanelli}, {Kent}
  et~al.}]{haynes18}
{Haynes} M.P., et~al., 2018, \apj, 861, 49

\bibitem[{{Hayward} et~al.(2012){Hayward}, {Jonsson}, {Kere{\v s}}
  et~al.}]{hayward12}
{Hayward} C.C., {Jonsson} P., {Kere{\v s}} D., {Magnelli} B., {Hernquist} L.,
  {Cox} T.J., 2012, \mnras, 424, 951

\bibitem[{{Hayward} et~al.(2014){Hayward}, {Torrey}, {Springel}, {Hernquist},
  \& {Vogelsberger}}]{hayward14}
{Hayward} C.C., {Torrey} P., {Springel} V., {Hernquist} L., {Vogelsberger} M.,
  2014, \mnras, 442, 1992

\bibitem[{{Hendry} et~al.(2005){Hendry}, {Smartt}, {Maund} et~al.}]{hendry05}
{Hendry} M.A., et~al., 2005, \mnras, 359, 906

\bibitem[{{Hjorth} \& {Bloom}(2012)}]{hjorthsn}
{Hjorth} J., {Bloom} J.S., 2012, Cambridge University Press, 169

\bibitem[{{Ho}(2019)}]{ho19rep}
{Ho} A., 2019, Transient Name Server Discovery Report, 1165

\bibitem[{{Hoogerwerf} et~al.(2001){Hoogerwerf}, {de Bruijne}, \& {de
  Zeeuw}}]{hoogerwerf01}
{Hoogerwerf} R., {de Bruijne} J.H.J., {de Zeeuw} P.T., 2001, \aap, 365, 49

\bibitem[{{Hopkins} et~al.(2006){Hopkins}, {Hernquist}, {Cox}
  et~al.}]{hopkins06b}
{Hopkins} P.F., {Hernquist} L., {Cox} T.J., {Di Matteo} T., {Robertson} B.,
  {Springel} V., 2006, \apjs, 163, 1

\bibitem[{{Huang} et~al.(2019){Huang}, {Shimoda}, {Urata} et~al.}]{huang19}
{Huang} K., et~al., 2019, \apjl, 878, L25

\bibitem[{{Joye} \& {Mandel}(2003)}]{ds9}
{Joye} W.A., {Mandel} E., 2003, In: {H.~E.~Payne, R.~I.~Jedrzejewski, \&
  R.~N.~Hook} (ed.) Astronomical Data Analysis Software and Systems XII, vol.
  295 of Astronomical Society of the Pacific Conference Series, 489

\bibitem[{{Kamphuis} \& {Briggs}(1992)}]{kamphuis92}
{Kamphuis} J., {Briggs} F., 1992, \aap, 253, 335

\bibitem[{{Karapetyan} et~al.(2018){Karapetyan}, {Hakobyan}, {Barkhudaryan}
  et~al.}]{karapetyan18}
{Karapetyan} A.G., {Hakobyan} A.A., {Barkhudaryan} L.V., {Mamon} G.A., {Kunth}
  D., {Adibekyan} V., {Turatto} M., 2018, \mnras, 481, 566

\bibitem[{{Kennicutt} et~al.(2011){Kennicutt}, {Calzetti}, {Aniano}
  et~al.}]{kingfish}
{Kennicutt} R.C., et~al., 2011, \pasp, 123, 1347

\bibitem[{{Kinugasa} et~al.(2002){Kinugasa}, {Kawakita}, {Ayani}
  et~al.}]{kinugasa02}
{Kinugasa} K., et~al., 2002, \apjl, 577, L97

\bibitem[{{Kippenhahn} \& {Weigert}(1990)}]{kippenhahn90}
{Kippenhahn} R., {Weigert} A., 1990, {Stellar Structure and Evolution},
  (Berlin:Springer)

\bibitem[{{Kuin} et~al.(2019){Kuin}, {Wu}, {Oates} et~al.}]{kuin19}
{Kuin} N.P.M., et~al., 2019, \mnras, 487, 2505

\bibitem[{{Leloudas} et~al.(2011){Leloudas}, {Gallazzi}, {Sollerman}
  et~al.}]{leloudas11}
{Leloudas} G., et~al., 2011, \aap, 530, A95

\bibitem[{{Leroy} et~al.(2009){Leroy}, {Walter}, {Bigiel} et~al.}]{leroy09}
{Leroy} A.K., et~al., 2009, \aj, 137, 4670

\bibitem[{{Liu} et~al.(2018){Liu}, {Zhang}, {Wang}, \& {Dai}}]{liu18}
{Liu} L.D., {Zhang} B., {Wang} L.J., {Dai} Z.G., 2018, \apjl, 868, L24

\bibitem[{{Lu} et~al.(1993){Lu}, {Hoffman}, {Groff}, {Roos}, \&
  {Lamphier}}]{lu93}
{Lu} N.Y., {Hoffman} G.L., {Groff} T., {Roos} T., {Lamphier} C., 1993, \apjs,
  88, 383

\bibitem[{{Lyutikov} \& {Toonen}(2019)}]{lyutikov19}
{Lyutikov} M., {Toonen} S., 2019, \mnras, 487, 5618

\bibitem[{{Marcum} et~al.(2001){Marcum}, {O'Connell}, {Fanelli}
  et~al.}]{marcum01}
{Marcum} P.M., et~al., 2001, \apjs, 132, 129

\bibitem[{{Margutti} et~al.(2019){Margutti}, {Metzger}, {Chornock}
  et~al.}]{margutti19}
{Margutti} R., et~al., 2019, \apj, 872, 18

\bibitem[{{Mauerhan} et~al.(2017){Mauerhan}, {Van Dyk}, {Johansson}
  et~al.}]{mauerhan17}
{Mauerhan} J.C., et~al., 2017, \apj, 834, 118

\bibitem[{{Maund}(2018)}]{maund18}
{Maund} J.R., 2018, \mnras, 476, 2629

\bibitem[{{Maund} \& {Smartt}(2009)}]{maund09}
{Maund} J.R., {Smartt} S.J., 2009, Science, 324, 486

\bibitem[{{Mazzali} et~al.(2002){Mazzali}, {Deng}, {Maeda} et~al.}]{mazzali02}
{Mazzali} P.A., et~al., 2002, \apjl, 572, L61

\bibitem[{{Micha{\l}owski} et~al.(2014){Micha{\l}owski}, {Hunt}, {Palazzi}
  et~al.}]{michalowski14}
{Micha{\l}owski} M.J., et~al., 2014, \aap, 562, A70

\bibitem[{{Micha{\l}owski} et~al.(2015){Micha{\l}owski}, {Gentile}, {Hjorth}
  et~al.}]{michalowski15hi}
{Micha{\l}owski} M.J., et~al., 2015, \aap, 582, A78

\bibitem[{{Micha{\l}owski} et~al.(2016){Micha{\l}owski}, {Castro Cer{\'o}n},
  {Wardlow} et~al.}]{michalowski16}
{Micha{\l}owski} M.J., et~al., 2016, \aap, 595, A72

\bibitem[{{Micha{\l}owski} et~al.(2018{\natexlab{a}}){Micha{\l}owski},
  {Gentile}, {Kr{\"u}hler} et~al.}]{michalowski18}
{Micha{\l}owski} M.J., et~al., 2018{\natexlab{a}}, \aap, 618, A104

\bibitem[{{Micha{\l}owski} et~al.(2018{\natexlab{b}}){Micha{\l}owski},
  {Karska}, {Rizzo} et~al.}]{michalowski18co}
{Micha{\l}owski} M.J., et~al., 2018{\natexlab{b}}, \aap, 617, A143

\bibitem[{{Micha{\l}owski} et~al.(2019){Micha{\l}owski}, {Kamphuis}, {Hjorth}
  et~al.}]{michalowski19}
{Micha{\l}owski} M.J., et~al., 2019, \aap, 627, A106

\bibitem[{{Mulcahy} et~al.(2017){Mulcahy}, {Beck}, \& {Heald}}]{mulcahy17}
{Mulcahy} D.D., {Beck} R., {Heald} G.H., 2017, \aap, 600, A6

\bibitem[{{Nakano} et~al.(2002){Nakano}, {Hirose}, {Kushida}, {Kushida}, \&
  {Li}}]{nakano12circ}
{Nakano} S., {Hirose} Y., {Kushida} R., {Kushida} Y., {Li} W., 2002, \iaucirc,
  7810

\bibitem[{{Oh} et~al.(2015){Oh}, {Kim}, \& {Lee}}]{oh16}
{Oh} S.H., {Kim} W.T., {Lee} H.M., 2015, \apj, 807, 73

\bibitem[{{Perley} et~al.(2017){Perley}, {Kr{\"u}hler}, {Schady}
  et~al.}]{perley17}
{Perley} D.A., et~al., 2017, \mnras, 465, L89

\bibitem[{{Perley} et~al.(2019){Perley}, {Mazzali}, {Yan} et~al.}]{perley19}
{Perley} D.A., et~al., 2019, \mnras, 484, 1031

\bibitem[{{Pettitt} et~al.(2016){Pettitt}, {Tasker}, \& {Wadsley}}]{pettitt16}
{Pettitt} A.R., {Tasker} E.J., {Wadsley} J.W., 2016, \mnras, 458, 3990

\bibitem[{{Pilbratt} et~al.(2010){Pilbratt}, {Riedinger}, {Passvogel}
  et~al.}]{herschel}
{Pilbratt} G.L., et~al., 2010, \aap, 518, L1

\bibitem[{{Poglitsch} et~al.(2010){Poglitsch}, {Waelkens}, {Geis}
  et~al.}]{pacs}
{Poglitsch} A., et~al., 2010, \aap, 518, L2

\bibitem[{{Prentice} et~al.(2018){Prentice}, {Maguire}, {Smartt}
  et~al.}]{prentice18}
{Prentice} S.J., et~al., 2018, \apjl, 865, L3

\bibitem[{{Rivera Sandoval} et~al.(2018){Rivera Sandoval}, {Maccarone}, {Corsi}
  et~al.}]{riverasandoval18}
{Rivera Sandoval} L.E., {Maccarone} T.J., {Corsi} A., {Brown} P.J., {Pooley}
  D., {Wheeler} J.C., 2018, \mnras, 480, L146

\bibitem[{{Roming} et~al.(2005){Roming}, {Kennedy}, {Mason} et~al.}]{uvot}
{Roming} P.W.A., et~al., 2005, \ssr, 120, 95

\bibitem[{{Roychowdhury} et~al.(2019){Roychowdhury}, {Arabsalmani}, \&
  {Kanekar}}]{roychowdhury19}
{Roychowdhury} S., {Arabsalmani} M., {Kanekar} N., 2019, \mnras, 485, L93

\bibitem[{{Scarano} \& {L{\'e}pine}(2013)}]{scarano13}
{Scarano} S., {L{\'e}pine} J.R.D., 2013, \mnras, 428, 625

\bibitem[{{Scheepmaker} et~al.(2009){Scheepmaker}, {Lamers}, {Anders}, \&
  {Larsen}}]{scheepmaker09}
{Scheepmaker} R.A., {Lamers} H.J.G.L.M., {Anders} P., {Larsen} S.S., 2009,
  \aap, 494, 81

\bibitem[{{Shu}(2016)}]{shu16}
{Shu} F.H., 2016, \araa, 54, 667

\bibitem[{{Smartt} et~al.(2002){Smartt}, {Vreeswijk}, {Ramirez-Ruiz}
  et~al.}]{smartt02}
{Smartt} S.J., {Vreeswijk} P.M., {Ramirez-Ruiz} E., {Gilmore} G.F., {Meikle}
  W.P.S., {Ferguson} A.M.N., {Knapen} J.H., 2002, \apjl, 572, L147

\bibitem[{{Smartt} et~al.(2004){Smartt}, {Maund}, {Hendry} et~al.}]{smartt04}
{Smartt} S.J., {Maund} J.R., {Hendry} M.A., {Tout} C.A., {Gilmore} G.F.,
  {Mattila} S., {Benn} C.R., 2004, Science, 303, 499

\bibitem[{{Soderberg} et~al.(2010){Soderberg}, {Chakraborti}, {Pignata}
  et~al.}]{soderberg10}
{Soderberg} A.M., et~al., 2010, \nat, 463, 513

\bibitem[{{Soker} et~al.(2019){Soker}, {Grichener}, \& {Gilkis}}]{soker19}
{Soker} N., {Grichener} A., {Gilkis} A., 2019, \mnras, 484, 4972

\bibitem[{{Stanway} et~al.(2015){Stanway}, {Levan}, {Tanvir}, {Wiersema}, \&
  {van der Laan}}]{stanway15}
{Stanway} E.R., {Levan} A.J., {Tanvir} N.R., {Wiersema} K., {van der Laan}
  T.P.R., 2015, \apjl, 798, L7

\bibitem[{{Th{\"o}ne} et~al.(2009){Th{\"o}ne}, {Micha{\l}owski}, {Leloudas}
  et~al.}]{thone09}
{Th{\"o}ne} C.C., {Micha{\l}owski} M.J., {Leloudas} G., {Cox} N.L.J., {Fynbo}
  J.P.U., {Sollerman} J., {Hjorth} J., {Vreeswijk} P.M., 2009, \apj, 698, 1307

\bibitem[{{Valenti} et~al.(2014){Valenti}, {Sand}, {Pastorello}
  et~al.}]{valenti14}
{Valenti} S., et~al., 2014, \mnras, 438, L101

\bibitem[{{Van Dyk} et~al.(2003){Van Dyk}, {Li}, \& {Filippenko}}]{vandyk03}
{Van Dyk} S.D., {Li} W., {Filippenko} A.V., 2003, \pasp, 115, 1289

\bibitem[{{Walter} et~al.(2008){Walter}, {Brinks}, {de Blok} et~al.}]{walter08}
{Walter} F., {Brinks} E., {de Blok} W.J.G., {Bigiel} F., {Kennicutt} R.C. Jr.,
  {Thornley} M.D., {Leroy} A., 2008, \aj, 136, 2563

\bibitem[{{Wang} et~al.(2003){Wang}, {Baade}, {H{\"o}flich}, \&
  {Wheeler}}]{wang03}
{Wang} L., {Baade} D., {H{\"o}flich} P., {Wheeler} J.C., 2003, \apj, 592, 457

\bibitem[{{Werner} et~al.(2004){Werner}, {Roellig}, {Low} et~al.}]{spitzer}
{Werner} M.W., et~al., 2004, \apjs, 154, 1

\bibitem[{{Wright}(2006)}]{wrightcalc}
{Wright} E.L., 2006, \pasp, 118, 1711

\bibitem[{{Zapartas} et~al.(2017){Zapartas}, {de Mink}, {Van Dyk}
  et~al.}]{zapartas17}
{Zapartas} E., et~al., 2017, \apj, 842, 125

\end{thebibliography}

\appendix

\section{Additional figure}

\begin{figure}
\begin{center}
\includegraphics[width=\szerkol,clip]{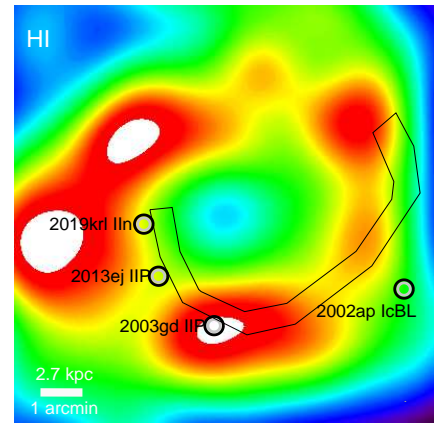} 
\end{center}
\caption{{\hi} VLA image (presented in Fig.~\ref{fig:image}) smoothed using a Gaussian with full width at half maximum of 100{\arcsec}. The spiral structure at this resolution resembles an irregular ring. The positions of SNe are marked by grey circles. The lines outline the main spiral arm. 
The panel is 10{\arcmin} per side. North is up and east is to the left.
}
\label{fig:smooth}
\end{figure}

\end{document}